\documentclass[11pt]{article}

\usepackage[T1]{fontenc}
\usepackage[utf8]{inputenc}

\usepackage[backend=biber, natbib=true]{biblatex}
\usepackage{microtype}

\usepackage{
  graphicx, booktabs, multirow, authblk,
  url, siunitx, amsmath, amssymb, bm,
  hyperref,bookmark
}

\usepackage[caption=false]{subfig}
\usepackage[font=small,labelfont=bf]{caption}
\usepackage[inline]{enumitem}
\usepackage[normalem]{ulem}
\usepackage{xcolor}
\usepackage{xspace}
\usepackage{chngpage}

\usepackage{tikz}
\usepackage{tikz-qtree}
\usetikzlibrary{positioning,arrows,calc,math,shapes.multipart,shapes.misc,chains,scopes}

\DeclareRobustCommand{\vect}[1]{\bm{#1}}
\newcommand{\changed}[1]{{#1}}
\newcommand{\mund}{\mathunderscore}
\newcommand{\G}{\mathcal{G}}
\newcommand{\rev}{\textsuperscript{r}}
\newcommand{\hasedget}{\texttt{has\mund{}edge($\G, u, v, t$)}\xspace} 
\newcommand{\hasedgei}{\texttt{has\mund{}edge($\G,\,u,\,v,\,t_{begin},\,t_{end}$)}\xspace} 
\newcommand{\nextacti}{\texttt{next\mund{}activation($\G,\,u,\,v,\,t_{begin},\,t_{end}$)}\xspace} 
\newcommand{\neighsi}{\texttt{neighbors($\G,\,u,\,t_{begin},\,t_{end}$)}\xspace} 
\newcommand{\rneighsi}{\texttt{neighbors\rev($\G,\,u,\,t_{begin},\,t_{end}$)}\xspace} 
\newcommand{\aggregi}{\texttt{aggregate($\G,\,t_{begin},\,t_{end}$)}\xspace} 
\newcommand{\actedgesi}{\texttt{activated\mund{}edges($\G,\,t_{begin},\,t_{end}$)}\xspace} 
\newcommand{\deactedgesi}{\texttt{deactivated\mund{}edges($\G,\,t_{begin},\,t_{end}$)}\xspace} 
\newcommand{\chgedgesi}{\texttt{changed\mund{}edges($\G,\,t_{begin},\,t_{end}$)}\xspace} 

\newcommand{\hasedgename}{\texttt{has\mund{}edge}\xspace} 
\newcommand{\neighsname}{\texttt{neighbors}\xspace} 
\newcommand{\rneighsname}{\texttt{neighbors\rev}\xspace} 
\newcommand{\aggregname}{\texttt{aggregate}\xspace} 

\title{A Review of In-Memory Space-Efficient Data Structures for Temporal Graphs}

\author[1]{Luiz F. A. Brito}

\author[1]{Bruno A. N. Travençolo}
\author[1]{Marcelo K. Albertini}

\affil[1]{Federal University of Uberlândia, Brazil}

\begin{document}

\maketitle
\begin{abstract}
Temporal graphs model relationships among entities over time. Recent studies applied temporal graphs to abstract complex systems such as continuous communication among participants of social networks. Often, the amount of data is larger than main memory, therefore, we need specialized structures that balance space usage and query efficiency. In this paper, we review space-efficient data structures that bring large temporal graphs from external memory to primary memory and speed up specialized queries. We found a great variety of studies using data compression techniques and self-indexed compressed data structures. We point further research directions to improve the current state-of-the-art.
\end{abstract}

\section{Introduction}\label{intro}

Widespread adoption of complex network concepts in information technologies has driven the creation of large volumes of data to be modelled as graphs.
This increasing volume of data produced at high speeds brings us new challenges.
On one hand, we need efficient computational mechanisms to persist data that evolve continuously over time on cheap external storage.
On the other hand, we need specialized techniques to load these data in faster (and more expensive) memories using minimal space and, then, process queries as fast as possible to construct valuable knowledge.

Temporal graphs serve as a modeling tool to represent phenomena that occurs over time~\citep{michail2016introduction}.
They describe complex systems as relationships among entities which often appear in the form of contacts or events defining when relationships begin or end.
Recent studies have applied temporal graphs to solve problems such as detecting community hierarchies~\citep{yang2011detecting} and predicting future events~\citep{liben2007link} during continuous communication among participants of social networks.

However, storing and querying large temporal graphs is difficult, especially when the amount of data grows unboundedly~\citep{caro2015data}.
Recently, space-efficient data structures have been developed to store and query temporal graphs in primary memory~\citep{caro2015data,caro2016compressed,brisaboa2018using}.
These data structures consider the theoretical optimal space to describe a data domain and  categorize data structures based on how much extra space they use to answer their queries.
For instance, the theoretical optimal space to represent a binary tree containing   $n$ nodes is $Z = \log_2{C_n}$ bits since the total amount of configurations in   this domain is $C_n$, the Catalan number~\citep{catalan}.
In this context, a data structure is \textit{implicit} if it spends $Z + O(1)$ bits, \textit{succinct} if it uses $Z + o(Z)$ bits, and \textit{compact} if it uses $Z + O(Z)$ bits~\citep{jacobson1988succinct}.

In this short review, we give an overview of space-efficient data structures that store temporal graphs in primary memory \changed{and support fundamental queries such as checking if there is a relationship active during a time interval, retrieving the neighborhood of an entity after a time interval, or aggregating in a conventional graph all snapshots inside a time interval.}
Other papers have also reviewed strategies to store or query graphs efficiently; however, they focused on different aspects.
For example, \citeauthor{shi2018graph} surveyed in \citeyear{shi2018graph} algorithms for graph processing in Graphic Processing Units (GPUs) to speed up queries;
\citeauthor{liu2018graph} revised in \citeyear{liu2018graph} summarization methods to extract the most important characteristics of graphs and, therefore, reduce query complexity;
\citeauthor{mccune2015thinking} investigated in \citeyear{mccune2015thinking} strategies to store and query graphs using a distributed setup;
\citeauthor{besta2018survey} reviewed in \citeyear{besta2018survey} compression techniques to manipulate static graphs in primary memory;
\citeauthor{historical-graphs} studied in \citeyear{historical-graphs} approaches to store and query sequences of static graphs;
and \changed{\citeauthor{overview-methods-evolving} in \citeyear{overview-methods-evolving} studied other approaches for temporal graphs whose focuses on query time.
The latter mentioned some data structures included in this review, however, the authors did not provided an in-depth description of the algorithms.
Other studies proposed special data structures used exclusively to answer high-level queries such as shortest path finding~\cite{shortestpath} and community detection~\cite{yang2011detecting}.
This type of data structure is not considered in this review.
}

\changed{Specifically,} the data structures included in this paper are: Time-interval Log per Edge (EdgeLog), which keeps a list of time intervals for each edge stored in adjacency lists~\citep{xuan2003computing}; Adjacency Log of Evelog, which stores edge events of activation and deactivation for each vertex~\citep{caro2015data}; Compact Adjacency Sequency (CAS), which stores compactly edge events of activation and deactivation~\citep{caro2015data}; Compact Events ordered by Time (CET), which stores edge events of activation and deactivation ordering by time~\citep{caro2015data}; Temporal Graph Compressed Suffix Array (TGCSA), which uses the Compressed Suffix Array (CSA)~\citep{csa} and transforms queries on temporal graph into matching operations in a string representation of contacts~\citep{brisaboa2018using}; and Compressed $k^d$ tree (c$k^d$-tree), which stores contacts in a compact representation of a $4$-dimensional array~\citep{caro2016compressed}.

We remark that each discussed data structure is appropriated for different scenarios.
The choice of the optimal data structure differs depending, for example, on the essential queries needed by the application.
For instance, an application that needs to retrieve frequent neighbors of an entity considering its incoming relationships at a given timestamp could not use EveLog and EdgeLog since they need to traverse the whole structure.
Differently, CAS, CET, TGCSA and c$k^d$-tree would be more suitable for this case.
Especially CET, TGCSA and c$k^d$-tree, that have the same time complexity to retrieve neighbors of an entity at a specific timestamp considering both its incoming or outgoing relationships.

Each data structure can also answer particular queries more adequately or spend less space depending on the characteristics of the temporal graph.
For example, CET can check if a relationship exists during a given interval in $O(\log{n})$ time, where $n$ is the number of vertices in the temporal graph, and, therefore, the time complexity is upper bounded solely by the number of vertices.
Differently, c$k^d$-tree can answer most operations with cost depending only on the number of contacts in the temporal graph. Thus, this data structure is suitable for temporal graphs with low ratio of contacts per vertices or edges.
In this review, we will present a more detailed discussion relating the data structures and the characteristics of the temporal graph.



The rest of this review is organized as follows.
In Section~\ref{sec:background} we introduce the background concepts about temporal graph, main strategies to index and query temporal graphs, and useful queries on temporal graphs.
In Section~\ref{sec:primary-memory} we review space-efficient data structures for primary memory.
We highlight self-indexed compression techniques, which reduce the space overhead needed and run low-level queries efficiently.
In Section~\ref{sec:discussion} we discuss several data structures regarding aspects of space usage and time complexity.
Finally, in Section~\ref{sec:conclusion} we give our perspective of current state of the structures and research possibilities.

\section{Background}\label{sec:background}

In this section, first we give formal definitions about temporal graphs.
Then, we introduce a set of queries commonly used on applications.
Finally, we briefly present general strategies to store temporal graphs efficiently trading space and efficiency when necessary.

\subsection{Temporal Graphs}
Temporal graphs model the changes of relationships among entities over time.
A temporal graph $\mathcal{G}$ is defined by a tuple $\mathcal{G} = (V, E, T, C)$, where a set $V$ of vertices represents entities; a set $E \in V \times V$ of edges between two vertices indicates a relationship; a set $T$ contains discrete timestamps; and a set $C \in E \times  T \times T$ of contacts $c = (e, t_{begin}, t_{end})$ represents activation and deactivation of edges.
By definition, given a contact $c=(e, t_{begin}, t_{end})$, edge $e$ is active during the interval $[t_{begin}, t_{end}]$, including $t_{begin}$ and $t_{end}$.

Temporal graphs are also often described as an ordered sequence $\mathcal{G} = G_1, G_2, \ldots, G_t$, where $G_t = (V_t, E_t)$ is a snapshot or, interchangeably, a conventional graph at time $t \in T$.
In this review we use these two notations to explain how techniques work.

\begin{figure}
    \centering
\begin{tikzpicture}[inner sep=1pt,text depth=.25ex,text height=1.75ex,font=\ttfamily]
\node[circle, draw=black, fill=black!15] (u0) at (0, 0) {a};
\node[circle, draw=black, fill=black!15] (v0) at (1, 0) {b};
\draw[->, thick] (u0) -- (v0);
\node[circle, draw=black, fill=black!15] (u1) at (0, -1) {a};
\node[circle, draw=black, fill=black!15] (v1) at (1, -1) {d};
\draw[->, thick] (u1) -- (v1);
\node[circle, draw=black, fill=black!15] (u2) at (0, -2) {b};
\node[circle, draw=black, fill=black!15] (v2) at (1, -2) {c};
\draw[->, thick] (u2) -- (v2);
\node[circle, draw=black, fill=black!15] (u3) at (0, -3) {b};
\node[circle, draw=black, fill=black!15] (v3) at (1, -3) {e};
\draw[->, thick] (u3) -- (v3);
\node[circle, draw=black, fill=black!15] (u4) at (0, -4) {d};
\node[circle, draw=black, fill=black!15] (v4) at (1, -4) {b};
\draw[->, thick] (u4) -- (v4);
\node[circle, draw=black, fill=black!15] (u5) at (0, -5) {e};
\node[circle, draw=black, fill=black!15] (v5) at (1, -5) {d};
\draw[->, thick] (u5) -- (v5);

\draw[dashed] (2+.7*0, .5) -- (2+.7*0, -7+1.5) node[below] {$0$};
\draw[dashed] (2+.7*1, .5) -- (2+.7*1, -7+1.5) node[below] {$1$};
\draw[dashed] (2+.7*2, .5) -- (2+.7*2, -7+1.5) node[below] {$2$};
\draw[dashed] (2+.7*3, .5) -- (2+.7*3, -7+1.5) node[below] {$3$};
\draw[dashed] (2+.7*4, .5) -- (2+.7*4, -7+1.5) node[below] {$4$};
\draw[dashed] (2+.7*5, .5) -- (2+.7*5, -7+1.5) node[below] {$5$};
\draw[dashed] (2+.7*6, .5) -- (2+.7*6, -7+1.5) node[below] {$6$};

\draw[very thick] (2+.7*1, 0) -- (2+.7*3, 0);
\draw[very thick] (2+.7*2, -1) -- (2+.7*3, -1);
\draw[very thick] (2+.7*4, -1) -- (2+.7*6, -1);
\draw[very thick] (2+.7*3, -2) -- (2+.7*6, -2);
\draw[very thick] (2+.7*3, -3) -- (2+.7*5, -3);
\draw[very thick] (2+.7*0, -4) -- (2+.7*6, -4);
\draw[very thick] (2+.7*3, -5) -- (2+.7*5, -5);
\end{tikzpicture}
\caption{Directed temporal graph $\mathcal{G} = (V, E, T, C)$ that has the set a vertices $V = \{a, b, c, d, e\}$, the set of edges $E = \{(a, b), (a, d), (b, c), (b, e), (d, b), (e, d)\}$, the lifetime within $T = [0, 6]$, and the set of contacts $C = \{((a, b), 1, 3),\allowbreak ((a, d), 2, 3),\allowbreak ((a, d), 4, 6),\allowbreak ((b, c), 3, 6),\allowbreak ((b, e), 3, 5),\allowbreak ((d, b), 0, 6),\allowbreak ((e, d), 3, 5)\}$.}\label{fig:temporal-graphs}
\end{figure}
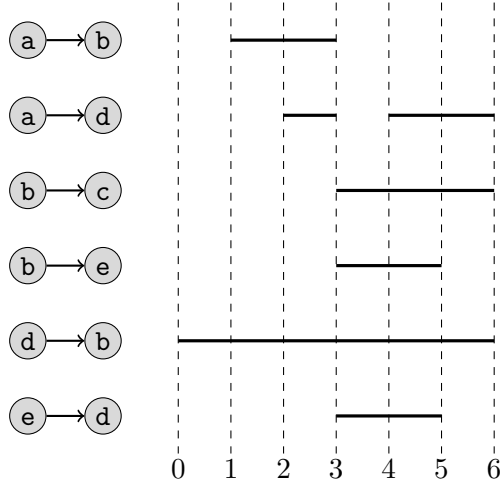

Temporal graphs are of two types: directed or undirected, depending on whether edges are ordered.
Directed edges express uni-directional edges as in followers-followees networks in which $(u, v) \neq (v, u)$.
Differently, undirected edges express bidirectionality as in collaboration networks in which authors co-write papers and $(u, v) = (v, u)$.
Figure~\ref{fig:temporal-graphs} illustrates a directed temporal graph.

\subsection{Queries on Temporal Graphs}

There are several queries to analyze temporal graphs.
Low-level queries solve smaller tasks such as checking if an edge is active at time $t$ or retrieving all neighbors of vertex $u$ at time $t$.
We can combine low-level queries to execute high-level ones, which
solve bigger problems such as finding the best temporal path connecting two vertices through time or clustering vertices along snapshots to detect community evolution.

\citeauthor{de2013compact} categorized in \citeyear{de2013compact} low-level queries in three types: edge, vertex, and time-related queries.
In all these types, we pass time-related arguments as input.
In the case we want to make a query about a specific timestamp we call it a \textit{point-time query}.
Instead, if we want to make a query on a range or interval we call it an \textit{interval query}.
Interval queries can also have two different semantics, \textit{weak} or \textit{strong}.
If an interval query has weak semantics, it is enough that a condition holds for snapshots of any moment during the interval.
Otherwise, if an interval query has strong semantics, the same condition needs to hold for all interval.
In this paper, if an interval $[t_{begin}, t_{end}]$ has $t_{begin} = t_{end}$, we consider it a point-time query and, thus, \textit{weak} and \textit{strong} semantics have the same effect.
Therefore, we will describe only \textit{interval queries} and, when necessary, we will explicitly distinguish it from \textit{point queries}.

Edge-related queries retrieve edge information during an interval $[t_{begin}, t_{end}]$.
For example, \changed{\hasedgei} checks if there is an edge $(u, v) \in E$ active during an interval $[t_{begin}, t_{end}]$.
If this query has weak semantics, it is enough to check whether a contact $\{(u, v), t'_{begin}, t'_{end}\}$ satisfies $t_{begin} \leq t'_{begin} \leq t_{end}$ or $t_{begin} \leq t'_{end} \leq t_{end}$
Instead, if it has strong semantics, we check whether the contact satisfies $t_{begin} \leq t'_{begin} < t'_{end} \leq t_{end}$.
Finally, $next\mathunderscore{}activation(\mathcal{G}, u, v, t)$ finds the next activation time of the edge $(u, v) \in E$ from timestamp $t$.
In order to retrieve the next activation, a contact must satisfy $t \leq t'_{begin} \leq t'_{end}$.
The answer for this query is an empty set if this $(u, v)$ has no further activation.
Also, different from others queries, $next\mathunderscore{}activation(\mathcal{G}, u, v, t)$ has support only for point-time semantics since passing an interval instead of a timestamp would not make sense to the purpose of this query.

Vertex-related queries retrieve vertices adjacent or incident to a given vertex during an interval $[t_{begin}, t_{end}]$.
The query $neighbors(\mathcal{G}, u, t_{begin}, t_{end})$ retrieves the vertices adjacent to $u$ that satisfy weak or strong semantics during the interval $[t_{begin}, t_{end}]$.
If it has weak semantics, it should retrieve all edges active at some timestamp in interval $[t_{begin}, t_{end}]$ by finding all pairs $(u, v) \in E$ such that $t_{begin} \leq t'_{begin} \leq  t_{end}$ or $t_{begin} \leq t'_{end} \leq  t_{end}$.

Instead, if it has strong semantics, it retrieves only vertices adjacent to $u$ active during all period of $[t_{begin}, t_{end}]$ by finding all pairs $(u, v) \in E$ such that $t_{begin} \leq t'_{begin} < t'_{end} \leq  t_{end}$.
Differently, $neighbors^{r}(\mathcal{G}, v, t_{begin}, t_{end})$ retrieves vertices incident to $v$ respecting weak or strong semantics during $[t_{begin}, t_{end}]$.

Time-related queries retrieve general graph information during an interval $[t_{begin}, t_{end}]$.
For example, the query $aggregate(\mathcal{G}, t_{begin}, t_{end})$ retrieves all edges $(u, v) \in E$ such that there is a contact $((u, v), t'_{begin}, t'_{end})$ that satisfies weak or strong semantics.
If $t = t_{begin} = t_{end}$, we say that this query extracts the snapshot $\mathcal{G}_t$ from $\mathcal{G}$ at timestamp $t$.
The query $activated\mathunderscore{}edges(\mathcal{G}, t_{begin}, t_{end})$ retrieves only edges $(u, v) \in E$ that has been activated during the interval $[t_{begin}, t_{end}]$.
This query has only weak semantics since $(u, v)$ cannot have been activated every timestamp in $[t_{begin}, t_{end}]$.

Similarly, the query $deactivated\mathunderscore{}edges(\mathcal{G}, t_{begin}, t_{end})$ retrieves only edges $(u, v)$ that has been deactivated during the interval $[t_{begin}, t_{end}]$ and has only weak semantics.
Finally, the query $changed\mathunderscore{}edges(\mathcal{G}, t_{begin}, t_{end})$ retrieves edges $(u, v)$ that have activation or deactivation events during the interval $[t_{begin}, t_{end}]$ and, differently, can have weak and strong semantics.
With strong semantics, this query would retrieve edges that intermittently change their connectivity status during the interval $[t_{begin}, t_{end}]$.

\begin{table}[]
    \caption{Examples of queries applied to the temporal graph shown in Figure~\ref{fig:temporal-graphs}.}\label{tab:queries}
\begin{adjustwidth}{-2in}{-2in}
    \centering
    \begin{tabular}{l|l|c|c|c} \toprule
        \textbf{Category} & \textbf{Query} & $\mathbf{t_{begin} = 2}\text{, }\mathbf{t_{end} = 2}$ & \multicolumn{2}{c}{$\mathbf{t_{begin} = 2}\text{, }\mathbf{t_{end} = 4}$} \\ \midrule
        & & \textbf{point-time query} & \multicolumn{2}{c}{\textbf{interval query}} \\
        & &  & \textbf{weak} & \textbf{strong} \\ \toprule
        \multirow{2}{*}{edge} & \changed{\hasedgei} & \textit{true} & \textit{true} & \textit{false} \\
                              & \changed{\nextacti} & $3$ & $\{\}$ & $\{\}$\\ \midrule
        \multirow{2}{*}{vertex} & \changed{\neighsi} & $\{ b \}$ & $\{ b \}$ & $\{ b \}$ \\
        & \changed{\rneighsi} & $\{a\}$ & $\{a, e \}$ & $\{\}$ \\ \midrule
        \multirow{4}{*}{time} & \changed{\aggregi} & $\{(a, b),(a, d),(d, b)\}$ & $\{(a, b), (a, d), (b, c), (b, e), (d, b), (e, d)\}$ & $\{(d, b)\}$ \\
                              & \changed{\actedgesi} & $\{(a, d)\}$ & $\{(a, d), (b, c), (b, e), (e, d)\}$ & - \\
                              & \changed{\deactedgesi} & $\{\}$ & $\{(a, b), (a, d)\}$ & -  \\
                              & \changed{\chgedgesi} & $\{(a, d)\}$ & $\{(a, b), (a, d), (b, c), (b, e), (e, d)\}$ & $\{(a, d)\}$ \\ \bottomrule
        \multicolumn{4}{c}{\changed{\footnotesize Symbol ``-'' means an invalid/non-possible query.}} \\
    \end{tabular}
\end{adjustwidth}
\end{table}

Table~\ref{tab:queries} illustrates the low-level queries using the temporal graph presented in Figure~\ref{fig:temporal-graphs}.
In the first column, we list the queries related to edge, vertex, and time; the second shows the results for point-time queries considering $t = 2$; and the third and fourth columns show the results for interval queries with weak and strong semantics considering the interval $[2, 4]$.
For instance, $neighbors^r(\mathcal{G}, d, 2)$ retrieves the set $\{(a, d)\}$ with a single entry since $(a, d)$ is the only edge active at timestamp $t = 2$.
Differently, the query $neighbors^r(\mathcal{G}, d, 2, 4)$ with weak semantics retrieves the set $\{(a, d), (e, d)\}$ in which edges $(a, d)$ and $(e, d)$ are active during interval $[2, 4]$.

Finally, the query $neighbors^r(\mathcal{G}, d, 2, 4)$ with strong semantics returns an empty set since there is no active incident edge to $d$ during the complete interval.

\section{Space-Efficient Data Structures for Querying Temporal Graphs in Primary Memory}\label{sec:primary-memory}

In this section, we present studies about space-efficient data structures to store and query large temporal graphs in primary memory.
These specialized data structures provide useful queries while spending little space as possible.
Some of them store a compressed version of data.
However, they compute queries without decompressing the data~\citep{brisaboa2014compressed,caro2015data,caro2016compressed,brisaboa2018using}.
The literature calls these approaches self-indexed space-efficient data structures.

For example, \citeauthor{grossi2003high} introduced in \citeyear{grossi2003high} a wavelet tree data structure that stores a sequence of $n$ symbols belonging to an alphabet of size $\sigma = |V| + |T|$ using only $n \lceil\log{\sigma}\rceil$ bits. The wavelet tree executes fundamental queries such as determining the frequency of symbols in a sub-range of the sequence in $O(\log{\sigma})$ time.
By using the wavelet tree, some data structures can quickly answer low-level queries for temporal graphs \changed{while retaining a minimal space}.

\changed{
  These data structures are different from previous works\footnote{See some examples of data structures for temporal graphs in~\cite{overview-methods-evolving}} in the sense that they are space-efficient and, at the same time, support queries with a cost often bounded by the worst-case complexities of standard data structures for the same purpose.
  For example, in~\cite{fvf}, the authors defined the FVF (Find, Verify, and Fix) framework.
  Although their framework answer queries quickly on average, it is not space efficient.
  In addition to all snapshots $\{G_1, G_2, \ldots, G_{\tau}\}$ of the temporal graph, this framework also stores representative graphs $G_{\cap}$ and $G_{\cup}$ containing the union and difference of snapshot clusters.
  In order to process a query, their algorithm first processes representative graphs in order to \emph{find} an initial solution, then it uses the individual snapshots to \emph{verify} the solution, and, if the solution does not apply, it try to \emph{fix} it.
  Another common approach stores only the differences between consecutive snapshots~\cite{deltas}.
  While this strategy is very space efficient, one might have to process the entire history in order to materialize a snapshot, thus the cost of simply checking whether there is a connection between two vertices at a given timestamp can be quite high.
  In order to minimize the query time when using differences, in~\cite{deltagraph}, the authors proposed a method that combines storing original snapshots and snapshot differences using a tree organization.
  Original snapshots are stored as leaves and snapshot differences as internal nodes.
  As in~\cite{fvf}, this strategy is fast, however, the space for storing snapshots and differences is high.
}

In the next subsections, we detail the following space-efficient data structures to index temporal graphs: time-interval log per edge (EdgeLog)~\citep{xuan2003computing}, adjacency log of events (EveLog)~\citep{caro2015data}, Compact Adjacency Sequence (CAS)~\citep{caro2015data}, Compressed Events ordered by Time (CET)~\citep{caro2015data}, Temporal Graph Compressed Suffix Array (TGCSA)~\citep{brisaboa2018using}, and Compressed $k^d$-tree~\citep{caro2016compressed}.
\changed{
  Most data structures presented in this paper is in the category of \emph{compact} data structures, which use $Z + O(Z)$ bits, where $Z$ is the optimal worst-case space to represent all configurations in the problem domain.
}
For clarity, in the following subsections we explain how the most fundamental queries work on these data structures after they are built.

\subsection{Time-interval Log per Edge}

\begin{figure}
    \centering
\begin{tikzpicture}[squarev/.style={draw,outer sep=0pt,inner sep=3pt,text depth=.25ex,text height=1.75ex,font=\ttfamily}, squaret/.style={draw,outer sep=0pt,inner sep=3pt,text depth=-0.25ex,text height=1.25ex,font=\ttfamily}]
\node[inner sep=5pt] (N@) {\normalsize $A$};
\node[anchor=north,draw,minimum height=3.2em,minimum width=1.5em,outer sep=0pt, fill=black!15](Na) at(N@.south) {}
 node[align=right,left] at(Na.west) {a};
\draw[->, semithick] ($(Na.east) + (-.5em, 0.8em)$) -- +(1em, 0)
 node (Ea@3) {};
\node[squarev, right] (Eab1) at (Ea@3.east) {b}
 node[squarev,right,fill=black!15] (Eab2) at (Eab1.east) {};
\draw[->, semithick] ($(Eab2.east) + (-.25em, 0)$) -- +(3.75em, 0)
 node (Eab3) {};
\draw[->, semithick] ($(Eab1.south) + (0, .25em)$) |- +(.5em, -1em)
 node (Tab@4) {};
\node[squaret,right] (Tab01) at (Tab@4.east) {$1$}
 node[squaret,right] (Tab02) at (Tab01.east) {$3$}
 node[squaret,right,fill=black!15] (Tab03) at (Tab02.east) {};
\draw[semithick] ($(Tab03.east) + (-.25em, 0.15em)$) -| +(.75em, -0.3em);
\draw[semithick] ($(Tab03.east) + (.25em, -0.25em)$) -- +(.5em, 0);
\draw[semithick] ($(Tab03.east) + (.35em, -0.45em)$) -- +(.3em, 0);
\node[squarev, right] (Ead1) at (Eab3.east) {d}
 node[squarev,right,fill=black!15] (Ead2) at (Ead1.east) {};
\draw[semithick] ($(Ead2.east) + (-.25em, 0.15em)$) -| +(0.75em, -0.3em);
\draw[semithick] ($(Ead2.east) + (.25em, -0.25em)$) -- +(.5em, 0);
\draw[semithick] ($(Ead2.east) + (.35em, -0.45em)$) -- +(.3em, 0);
\draw[->, semithick] ($(Ead1.south) + (0, .25em)$) |- +(.5em, -1em)
 node (Tad@4) {};
\node[squaret,right] (Tad01) at (Tad@4.east) {$2$}
 node[squaret,right] (Tad02) at (Tad01.east) {$3$}
 node[squaret,right,fill=black!15] (Tad03) at (Tad02.east) {};
\draw[->, semithick] ($(Tad03.east) + (-.25em, 0)$) -- +(.75em, 0)
 node (Tad04) {};
\node[squaret,right] (Tad11) at (Tad04.east) {$4$}
 node[squaret,right] (Tad12) at (Tad11.east) {$6$}
 node[squaret,right,fill=black!15] (Tad13) at (Tad12.east) {};
\draw[semithick] ($(Tad13.east) + (-.25em, 0.15em)$) -| +(.75em, -0.3em);
\draw[semithick] ($(Tad13.east) + (.25em, -0.25em)$) -- +(.5em, 0);
\draw[semithick] ($(Tad13.east) + (.35em, -0.45em)$) -- +(.3em, 0);

\node[anchor=north,draw,minimum height=3.2em,minimum width=1.5em,outer sep=0pt, fill=black!15](Nb) at(Na.south) {}
 node[align=right,left] at(Nb.west) {b};
\draw[->, semithick] ($(Nb.east) + (-.5em, 0.8em)$) -- +(1em, 0)
 node (Eb@3) {};
\node[squarev, right] (Ebc1) at (Eb@3.east) {c}
 node[squarev,right,fill=black!15] (Ebc2) at (Ebc1.east) {};
\draw[->, semithick] ($(Ebc2.east) + (-.25em, 0)$) -- +(3.75em, 0)
 node (Ebc3) {};
\draw[->, semithick] ($(Ebc1.south) + (0, .25em)$) |- +(.5em, -1em)
 node (Tbc@4) {};
\node[squaret,right] (Tbc01) at (Tbc@4.east) {$3$}
 node[squaret,right] (Tbc02) at (Tbc01.east) {$6$}
 node[squaret,right,fill=black!15] (Tbc03) at (Tbc02.east) {};
\draw[semithick] ($(Tbc03.east) + (-.25em, 0.15em)$) -| +(.75em, -0.3em);
\draw[semithick] ($(Tbc03.east) + (.25em, -0.25em)$) -- +(.5em, 0);
\draw[semithick] ($(Tbc03.east) + (.35em, -0.45em)$) -- +(.3em, 0);
\node[squarev, right] (Ebe1) at (Ebc3.east) {e}
 node[squarev,right,fill=black!15] (Ebe2) at (Ebe1.east) {};
\draw[semithick] ($(Ebe2.east) + (-.25em, 0.15em)$) -| +(0.75em, -0.3em);
\draw[semithick] ($(Ebe2.east) + (.25em, -0.25em)$) -- +(.5em, 0);
\draw[semithick] ($(Ebe2.east) + (.35em, -0.45em)$) -- +(.3em, 0);
\draw[->, semithick] ($(Ebe1.south) + (0, .25em)$) |- +(.5em, -1em)
 node (Tbe@4) {};
\node[squaret,right] (Tbe01) at (Tbe@4.east) {$3$}
 node[squaret,right] (Tbe02) at (Tbe01.east) {$5$}
 node[squaret,right,fill=black!15] (Tbe03) at (Tbe02.east) {};
\draw[semithick] ($(Tbe03.east) + (-.25em, 0.15em)$) -| +(.75em, -0.3em);
\draw[semithick] ($(Tbe03.east) + (.25em, -0.25em)$) -- +(.5em, 0);
\draw[semithick] ($(Tbe03.east) + (.35em, -0.45em)$) -- +(.3em, 0);

\node[anchor=north,draw,minimum height=3.2em,minimum width=1.5em,outer sep=0pt, fill=black!15](Nc) at(Nb.south) {}
 node[align=right,left] at(Nc.west) {c};
\draw[semithick] ($(Nc.east) + (-.5em, 0.8em)$) -| +(1em, -0.3em);
\draw[semithick] ($(Nc.east) + (.25em, 0.4em)$) -- +(.5em, 0);
\draw[semithick] ($(Nc.east) + (.35em, 0.2em)$) -- +(.3em, 0);

\node[anchor=north,draw,minimum height=3.2em,minimum width=1.5em,outer sep=0pt, fill=black!15](Nd) at(Nc.south) {}
 node[align=right,left] at(Nd.west) {d};
\draw[->, semithick] ($(Nd.east) + (-.5em, 0.8em)$) -- +(1em, 0)
 node (Ed@3) {};
\node[squarev, right] (Edb1) at (Ed@3.east) {b}
 node[squarev,right,fill=black!15] (Edb2) at (Edb1.east) {};
\draw[semithick] ($(Edb2.east) + (-.25em, 0.15em)$) -| +(0.75em, -0.3em);
\draw[semithick] ($(Edb2.east) + (.25em, -0.25em)$) -- +(.5em, 0);
\draw[semithick] ($(Edb2.east) + (.35em, -0.45em)$) -- +(.3em, 0);
\draw[->, semithick] ($(Edb1.south) + (0, .25em)$) |- +(.5em, -1em)
 node (Tdb@4) {};
\node[squaret,right] (Tdb01) at (Tdb@4.east) {$0$}
 node[squaret,right] (Tdb02) at (Tdb01.east) {$6$}
 node[squaret,right,fill=black!15] (Tdb03) at (Tdb02.east) {};
\draw[semithick] ($(Tdb03.east) + (-.25em, 0.15em)$) -| +(.75em, -0.3em);
\draw[semithick] ($(Tdb03.east) + (.25em, -0.25em)$) -- +(.5em, 0);
\draw[semithick] ($(Tdb03.east) + (.35em, -0.45em)$) -- +(.3em, 0);

\node[anchor=north,draw,minimum height=3.2em,minimum width=1.5em,outer sep=0pt, fill=black!15](Ne) at(Nd.south) {}
 node[align=right,left] at(Ne.west) {e};
\draw[->, semithick] ($(Ne.east) + (-.5em, 0.8em)$) -- +(1em, 0)
 node (Ee@3) {};
\node[squarev, right] (Eed1) at (Ee@3.east) {d}
 node[squarev,right,fill=black!15] (Eed2) at (Eed1.east) {};
\draw[semithick] ($(Eed2.east) + (-.25em, 0.15em)$) -| +(0.75em, -0.3em);
\draw[semithick] ($(Eed2.east) + (.25em, -0.25em)$) -- +(.5em, 0);
\draw[semithick] ($(Eed2.east) + (.35em, -0.45em)$) -- +(.3em, 0);
\draw[->, semithick] ($(Eed1.south) + (0, .25em)$) |- +(.5em, -1em)
 node (Ted@4) {};
\node[squaret,right] (Ted01) at (Ted@4.east) {$3$}
 node[squaret,right] (Ted02) at (Ted01.east) {$5$}
 node[squaret,right,fill=black!15] (Ted03) at (Ted02.east) {};
\draw[semithick] ($(Ted03.east) + (-.25em, 0.15em)$) -| +(.75em, -0.3em);
\draw[semithick] ($(Ted03.east) + (.25em, -0.25em)$) -- +(.5em, 0);
\draw[semithick] ($(Ted03.east) + (.35em, -0.45em)$) -- +(.3em, 0);

\end{tikzpicture}
    \caption{EdgeLog representation of the temporal graph shown in Figure~\ref{fig:temporal-graphs}. EdgeLog structure stores an array $A$ containing adjacency lists indexed by source vertex $u$. Each adjacency list $A[u]$ has target vertices $v$. For each vertex $v$ in $A[u]$, there is a list of intervals during which the edge $(u, v)$ is active. EdgeLog compresses these lists with DeltaGap and stores them in a contiguous space, instead of a list structure with pointers.}\label{fig:edgelog}
\end{figure}
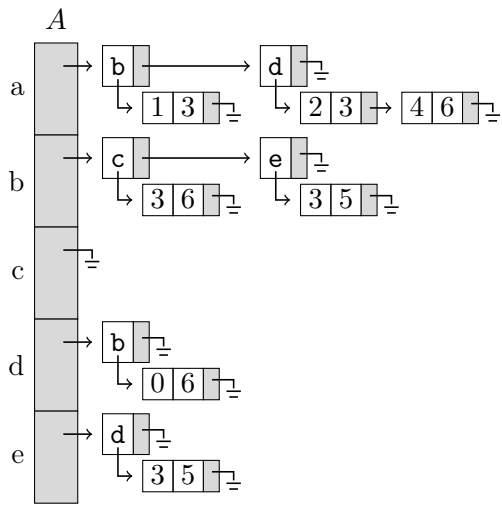

Time-interval Log per Edge (EdgeLog) keeps a list of time intervals for each edge~\citep{xuan2003computing}\footnote{Source code is available at \url{https://github.com/diegocaro/edgelogbase/}.}.
As shown in Figure~\ref{fig:edgelog}, this structure organizes contacts similarly to an inverted index~\citep{anh2005inverted} with three levels.
In the base level, it has an array $A$ indexed by source vertices $u$ containing pointers to adjacency lists.
In the first level, each adjacency list stores target vertices $v$ and another list containing, in the second level, the time intervals in which edges $(u, v)$ are active.
In order to cut the space overhead, it compresses both adjacency and interval lists.
First, each list is kept sorted to speed up queries by using binary search and to represent sequences of vertex and timestamps identifiers as DeltaGap encodings~\citep{adler2001towards}.
This encoding transforms sequences of numbers $L = l_1, l_2, \ldots, l_n$ on differences of subsequent numbers in the form $D = l_2 - l_1, l_3 - l_2, \ldots, l_n - l_{n-1}$.
Then, variable bit-wise compression techniques, such as Huffman Code \citep{knuth1985dynamic}, or word-wise compression techniques, such as PForDelta \citep{zukowski2006super} are applied to compress DeltaGap encodings.

This strategy reduces space usage because $D$ has lower entropy than $L$ and represents resulting values with fewer bits due to the lower range of the numbers. For commonly used encoding techniques for graphs we suggest the survey by \citeauthor{besta2018survey} and the textbook by~\citeauthor{navarro2016compact} published, respectively, in \citeyear{besta2018survey} and \citeyear{navarro2016compact}.

\subsubsection{Operation $\vect{has\mathunderscore{}edge}$ in EdgeLog}

In EdgeLog, an algorithm to answer \changed{\hasedgei} first decompresses the adjacency list $A[u]$ into a contiguous space in   memory, then it performs a binary search to find $v$ and the corresponding compressed list of intervals $T_v$.
Next, it decompresses $T_v$ and checks if the interval $[t_{begin}, t_{end}]$ overlaps with some interval in $T_v$ by performing a second binary search.
If so, an edge $(u, v)$ exists during $[t_{begin}, t_{end}]$.

\subsubsection{Operation $\vect{neighbors}$ in EdgeLog}
Similarly to the \changed{\hasedgename} operation, an algorithm to answer \changed{\neighsi} iterates all target vertices $v'$ in $A[u]$ and checks if the list of intervals $T_{v'}$ correspond to interval $[t_{begin}, t_{end}]$, according to weak or strong semantics.
If the query has weak semantics, then it checks if $[t_{begin}, t_{end}]$ overlaps with some $[t'_{begin}, t'_{end}] \in T_v'$.
Otherwise, if it has strong semantics, then the algorithm checks if there is some interval in $T_v'$ such that $t'_{begin} \leq t_{begin} < t_{end} \leq t'_{end}$.
The result is a list containing all neighbors that satisfy these conditions.

\subsubsection{Operation $\vect{neighbors^r}$ in EdgeLog}
EdgeLog indexes edges only by source vertices, therefore, there is no efficient way to compute reverse queries such as \changed{\rneighsi}.

A naïve algorithm decompresses every list in first level and test the interval conditions on every $(v, T_{v}) \in A[u']$ for all $u' \in  V$.
Another approach is to keep a second EdgeLog structure that indexes edges by target vertices instead.
Therefore, an algorithm to answer \changed{\rneighsi} runs \changed{\neighsi} on this second structure.
However, this approach doubles the space required to store temporal graphs.

\subsection{Adjacency Log of Events}

 \begin{figure}
     \centering
\begin{tikzpicture}[square/.style={draw,outer sep=0pt,inner sep=3pt,text depth=.25ex,text height=1.75ex,font=\ttfamily}]
\node[inner sep=5pt] (N@) {\normalsize $E$};
\node[anchor=north,draw,minimum height=2em,minimum width=1.5em,outer sep=0pt, fill=black!15](Na) at (N@.south) {}
 node[align=right,left] at(Na.west) {a};
\draw[->, semithick] ($(Na.east) + (-.5em, 0)$) -- +(1em, 0)
 node (Ea@4) {};
\node[square, right] (Ea01) at (Ea@4.east) {b}
 node[square, right] (Ea02) at (Ea01.east) {$1$}
 node[square, right, fill=black!15] (Ea03) at (Ea02.east) {};
\draw[->, semithick] ($(Ea03.east) + (-.25em, 0)$) -- +(0.75em, 0)
 node (Ea04) {};
\node[square, right] (Ea11) at (Ea04.east) {d}
 node[square, right] (Ea12) at (Ea11.east) {$2$}
 node[square, right, fill=black!15] (Ea13) at (Ea12.east) {};
\draw[->, semithick] ($(Ea13.east) + (-.25em, 0)$) -- +(0.75em, 0)
 node (Ea14) {};
\node[square, right] (Ea21) at (Ea14.east) {b}
 node[square, right] (Ea22) at (Ea21.east) {$3$}
 node[square, right, fill=black!15] (Ea23) at (Ea22.east) {};
\draw[->, semithick] ($(Ea23.east) + (-.25em, 0)$) -- +(0.75em, 0)
 node (Ea24) {};
\node[square, right] (Ea31) at (Ea24.east) {d}
 node[square, right] (Ea32) at (Ea31.east) {$3$}
 node[square, right, fill=black!15] (Ea33) at (Ea32.east) {};
\draw[->, semithick] ($(Ea33.east) + (-.25em, 0)$) -- +(0.75em, 0)
 node (Ea34) {};
\node[square, right] (Ea41) at (Ea34.east) {d}
 node[square, right] (Ea42) at (Ea41.east) {$4$}
 node[square, right, fill=black!15] (Ea43) at (Ea42.east) {};
\draw[->, semithick] ($(Ea43.east) + (-.25em, 0)$) -- +(0.75em, 0)
 node (Ea44) {};
\node[square, right] (Ea51) at (Ea44.east) {d}
 node[square, right] (Ea52) at (Ea51.east) {$6$}
 node[square, right, fill=black!15] (Ea53) at (Ea52.east) {};
\draw[semithick] ($(Ea53.east) + (-.25em, 0.15em)$) -| +(0.75em, -0.3em);
\draw[semithick] ($(Ea53.east) + (.25em, -0.25em)$) -- +(.5em, 0);
\draw[semithick] ($(Ea53.east) + (.35em, -0.45em)$) -- +(.3em, 0);

\node[anchor=north,draw,minimum height=2em,minimum width=1.5em,outer sep=0pt, fill=black!15](Nb) at (Na.south) {}
 node[align=right,left] at(Nb.west) {b};
\draw[->, semithick] ($(Nb.east) + (-.5em, 0)$) -- +(1em, 0)
 node (Eb@4) {};
\node[square, right] (Eb01) at (Eb@4.east) {c}
 node[square, right] (Eb02) at (Eb01.east) {$3$}
 node[square, right, fill=black!15] (Eb03) at (Eb02.east) {};
\draw[->, semithick] ($(Eb03.east) + (-.25em, 0)$) -- +(0.75em, 0)
 node (Eb04) {};
\node[square, right] (Eb11) at (Eb04.east) {e}
 node[square, right] (Eb12) at (Eb11.east) {$3$}
 node[square, right, fill=black!15] (Eb13) at (Eb12.east) {};
\draw[->, semithick] ($(Eb13.east) + (-.25em, 0)$) -- +(0.75em, 0)
 node (Eb14) {};
\node[square, right] (Eb21) at (Eb14.east) {e}
 node[square, right] (Eb22) at (Eb21.east) {$5$}
 node[square, right, fill=black!15] (Eb23) at (Eb22.east) {};
\draw[->, semithick] ($(Eb23.east) + (-.25em, 0)$) -- +(0.75em, 0)
 node (Eb24) {};
\node[square, right] (Eb31) at (Eb24.east) {c}
 node[square, right] (Eb32) at (Eb31.east) {$6$}
 node[square, right, fill=black!15] (Eb33) at (Eb32.east) {};
\draw[semithick] ($(Eb33.east) + (-.25em, 0.15em)$) -| +(0.75em, -0.3em);
\draw[semithick] ($(Eb33.east) + (.25em, -0.25em)$) -- +(.5em, 0);
\draw[semithick] ($(Eb33.east) + (.35em, -0.45em)$) -- +(.3em, 0);

\node[anchor=north,draw,minimum height=2em,minimum width=1.5em,outer sep=0pt, fill=black!15](Nc) at (Nb.south) {}
 node[align=right,left] at(Nc.west) {c};
\draw[semithick] ($(Nc.east) + (-.5em, 0.15em)$) -| +(1em, -0.3em);
\draw[semithick] ($(Nc.east) + (.25em, -0.25em)$) -- +(.5em, 0);
\draw[semithick] ($(Nc.east) + (.35em, -0.45em)$) -- +(.3em, 0);

\node[anchor=north,draw,minimum height=2em,minimum width=1.5em,outer sep=0pt, fill=black!15](Nd) at (Nc.south) {}
 node[align=right,left] at(Nd.west) {d};
\draw[->, semithick] ($(Nd.east) + (-.5em, 0)$) -- +(1em, 0)
 node (Ed@4) {};
\node[square, right] (Ed01) at (Ed@4.east) {b}
 node[square, right] (Ed02) at (Ed01.east) {$0$}
 node[square, right, fill=black!15] (Ed03) at (Ed02.east) {};
\draw[->, semithick] ($(Ed03.east) + (-.25em, 0)$) -- +(0.75em, 0)
 node (Ed04) {};
\node[square, right] (Ed11) at (Ed04.east) {b}
 node[square, right] (Ed12) at (Ed11.east) {$6$}
 node[square, right, fill=black!15] (Ed13) at (Ed12.east) {};
\draw[semithick] ($(Ed13.east) + (-.25em, 0.15em)$) -| +(0.75em, -0.3em);
\draw[semithick] ($(Ed13.east) + (.25em, -0.25em)$) -- +(.5em, 0);
\draw[semithick] ($(Ed13.east) + (.35em, -0.45em)$) -- +(.3em, 0);

\node[anchor=north,draw,minimum height=2em,minimum width=1.5em,outer sep=0pt, fill=black!15](Ne) at (Nd.south) {}
 node[align=right,left] at(Ne.west) {e};
\draw[->, semithick] ($(Ne.east) + (-.5em, 0)$) -- +(1em, 0)
 node (Ee@4) {};
\node[square, right] (Ee01) at (Ee@4.east) {d}
 node[square, right] (Ee02) at (Ee01.east) {$3$}
 node[square, right, fill=black!15] (Ee03) at (Ee02.east) {};
\draw[->, semithick] ($(Ee03.east) + (-.25em, 0)$) -- +(0.75em, 0)
 node (Ee04) {};
\node[square, right] (Ee11) at (Ee04.east) {d}
 node[square, right] (Ee12) at (Ee11.east) {$5$}
 node[square, right, fill=black!15] (Ee13) at (Ee12.east) {};
\draw[semithick] ($(Ee13.east) + (-.25em, 0.15em)$) -| +(0.75em, -0.3em);
\draw[semithick] ($(Ee13.east) + (.25em, -0.25em)$) -- +(.5em, 0);
\draw[semithick] ($(Ee13.east) + (.35em, -0.45em)$) -- +(.3em, 0);

\end{tikzpicture}
     \caption{EveLog representation of the temporal graph shown in Figure~\ref{fig:temporal-graphs}.
This structure has an array $E$ containing event lists indexed by source vertices $u$. Each event list $E[u]$ has pairs $(v, t)$ that represent events of activation or deactivation of edge $(u, v)$ at time $t$.
At implementation level, EveLog separates $E[u]$ in two lists, $V$ and $T$, containing target vertices and timestamps.
It compresses the ordered $T$ lists using DeltaGap and $V$ lists using ETDC{}.
Then it stores them in a contiguous space instead of a list structure with pointers. Note that we can check if $(u, v)$ is active at timestamp $t$ by determining the frequency of $v$ symbols in $E[u]$ until the last $t$ symbol.}\label{fig:evelog}
\end{figure}
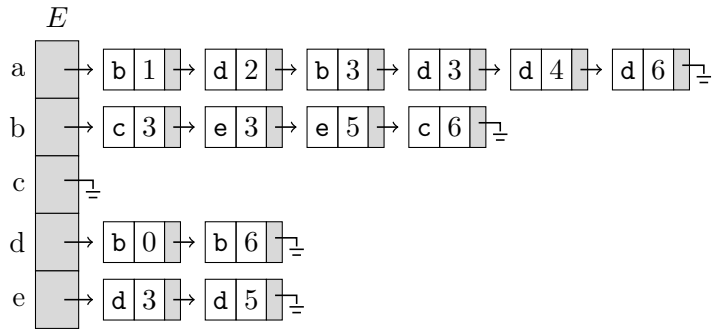

Adjacency Log of Events (EveLog) data structure stores events about edge activation and deactivation~\citep{caro2015data}\footnote{Source code is available at \url{https://github.com/diegocaro/adjlogbase/}.}.
As shown in Figure~\ref{fig:evelog}, EveLog has an array $E$ indexed by source      vertices $u$ containing pointers to lists of events ordered by time.
Each list has events $(v, t) \in V \times T$ that represent activation or      deactivation of edge $(u, v)$ at time $t$.
EveLog does not store explicitly whether events represent activation or             deactivation. Instead, it uses a parity property~\citep{caro2015data}.
As $E[u]$ is time-ordered, we need only to count the number of occurrences of    target vertices $v$ until some timestamp $t$, if it is odd, then edge $(u, v)$ is   active at $t$, otherwise, it is not.

Each list of events is compressed to reduce space.
First, EveLog separates lists $E[u]$ containing elements $(v, t)$ in two different lists $\mathcal{V} = v_1, v_2, \ldots, v_k$ and $\mathcal{T} = t_1, t_2, \ldots, t_k$.
As $\mathcal{T}$ is ordered, EveLog applies DeltaGap compression, the same approach used to compress EdgeLog lists.
However, $\mathcal{V}$ is not ordered and EveLog cannot sort separately, otherwise, EveLog would lose the mapping between pairs on $\mathcal{V}$ and $\mathcal{T}$.
Therefore, DeltaGap encodings cannot be used to decrease the entropy of $\mathcal{V}$ lists.
\citeauthor{caro2015data} chose in \citeyear{caro2015data} the End-Tagged Dense Codes (ETDC)~\citep{brisaboa2003efficient} technique, often used in information retrieval context, to compress $\mathcal{V}$ lists since it is faster than Huffman Code while producing compressed sequences only $2.5\%$ bigger.

\subsubsection{Operation $\vect{has\mathunderscore{}edge}$ in EveLog}
An algorithm to answer \changed{\hasedget} first decompresses both lists $\mathcal{V}$ and $\mathcal{T}$ associated with $E[u]$.
Then, it performs binary searches in $\mathcal{T}$ to find the positions $i$ and $j$ associated with the last timestamp symbols of $t_{begin}$ and $t_{end}$, respectively.
Next, it finds the frequencies $f_{begin}$ and $f_{end}$ of elements $v$ inside the subsequences $\mathcal{V}[1, i] = v_1, v_2, \ldots, v_i$ and $\mathcal{V}[i , j] = v_i, v_{i + 1}, \ldots, v_j$.
If $f_{begin}$ is odd, then edge $(u, v)$ is active at time $t_{begin}$, and if $f_{end}$ is greater than zero, then $(u, v)$ is active at some timestamp during the interval $[t_{begin}, t_{end}]$, otherwise is not active.
Note that, if it is a point-based query, $t = t_{begin} = t_{end}$ and $f = f_{begin} = f_{end}$, then the algorithm only needs to check if $f$ is odd or even to answer whether $(u, v)$ is, respectively, active or not at timestamp $t$.

\subsubsection{Operation $\vect{neighbors}$ in EveLog}

An algorithm to answer \changed{\neighsi} performs a   similar approach, however, it counts frequencies for each possible vertex $v'$.
First, it finds the frequencies $f_{begin}$ and $f_{end}$ for all vertices $v'$   in both subsequences $\mathcal{V}[1, i]$ and $\mathcal{V}[i, j]$, where $i$ and $j$ are positions in $\mathcal{T}$ associated with times $t_{begin}$ and $t_{end}$,     respectively.
Next, depending on the query, it checks weak or strong semantics.
If $neighbors(\mathcal{G}, u, t_{begin}, t_{end})$ has weak semantics, then the      algorithm retrieves all contacts in which edge $(u, v')$ is active at any timestamp during $[t_{begin}, t_{end}]$.
Therefore, it returns a contact if edge $(u, v')$ enters active in $[t_{begin}, t_{end}]$ --- same as $f_{begin}$ being odd --- or if it enters inactive and,       later, it activates at some timestamp during $[t_{begin}, t_{end}]$  --- same as $f_{end}$ being greater than $0$.
Instead, if the query has strong semantics, then it only retrieves contacts in which edge $(u, v')$ is active during the interval $[t_{begin}, t_{end}]$.
Therefore, it returns a contact if edge $(u, v')$ enters active  --- same as    $f_{begin}$ being odd --- and keeps active until $t_{end}$ --- same as $f_{end}$ being equals to $0$.

\subsubsection{Operation $\vect{neighbors^r}$ in EveLog}

EveLog cannot answer \changed{\rneighsi} efficiently.
An algorithm would decompress every event list and, for each one, it would count frequencies of edge events linearly until time $t$.
As in EdgeLog, a workaround is to build a second EveLog structure and use it to run the direct query instead.
However, as EveLog stores every event, the negative impact of doubling the space is higher than for the EdgeLog structure.

\subsection{Compact Adjacency Sequence (CAS)}

\begin{figure}
    \centering
\begin{tikzpicture}[square/.style={draw,outer sep=0pt,inner sep=3pt,text depth=.25ex,text height=1.75ex,font=\ttfamily}]
\node[square,text=red] (S0) {1}
 node[square, right,text=black] (S1) at (S0.east) {b}
 node[square, right,text=red] (S2) at (S1.east) {2}
 node[square, right,text=black] (S3) at (S2.east) {d}
 node[square, right,text=red] (S4) at (S3.east) {3}
 node[square, right,text=black] (S5) at (S4.east) {b}
 node[square, right,text=black] (S6) at (S5.east) {d}
 node[square, right,text=red] (S7) at (S6.east) {4}
 node[square, right,text=black] (S8) at (S7.east) {d}
 node[square, right,text=red] (S9) at (S8.east) {6}
 node[square, right,text=black] (S10) at (S9.east) {d}
 node[square, right,text=red] (S11) at (S10.east) {3}
 node[square, right,text=black] (S12) at (S11.east) {c}
 node[square, right,text=black] (S13) at (S12.east) {e}
 node[square, right,text=red] (S14) at (S13.east) {5}
 node[square, right,text=black] (S15) at (S14.east) {e}
 node[square, right,text=red] (S16) at (S15.east) {6}
 node[square, right,text=black] (S17) at (S16.east) {c}
 node[square, right,text=red] (S18) at (S17.east) {0}
 node[square, right,text=black] (S19) at (S18.east) {b}
 node[square, right,text=red] (S20) at (S19.east) {6}
 node[square, right,text=black] (S21) at (S20.east) {b}
 node[square, right,text=red] (S22) at (S21.east) {3}
 node[square, right,text=black] (S23) at (S22.east) {d}
 node[square, right,text=red] (S24) at (S23.east) {5}
 node[square, right,text=black] (S25) at (S24.east) {d};
\node[left= 0.3em of S0] {\normalsize $S$};
\node[square, anchor=north,inner sep=2.2pt,font=\ttfamily\small,yshift=-1em] (B13) at (S11.south) {0};
\draw[->,semithick] (B13.north) -- (S11.south);
\node[square, anchor=west,inner sep=2pt,inner sep=2.2pt,font=\ttfamily\small] (B14) at (B13.east) {0};
\node[square, anchor=west,inner sep=2pt,inner sep=2.2pt,font=\ttfamily\small] (B15) at (B14.east) {0};
\node[square, anchor=west,inner sep=2pt,inner sep=2.2pt,font=\ttfamily\small] (B16) at (B15.east) {0};
\node[square, anchor=west,inner sep=2pt,inner sep=2.2pt,font=\ttfamily\small] (B17) at (B16.east) {0};
\node[square, anchor=west,inner sep=2pt,inner sep=2.2pt,font=\ttfamily\small] (B18) at (B17.east) {0};
\node[square, anchor=west,inner sep=2pt,inner sep=2.2pt,font=\ttfamily\small] (B19) at (B18.east) {0};
\node[square, anchor=west,inner sep=2pt,inner sep=2.2pt,font=\ttfamily\small] (B20) at (B19.east) {1};
\node[circle,draw,fill=black!15,inner sep=1pt,minimum size=1.1em,below= 1.5mm of B20] {$c$};
\node[cross out,inner sep=1pt,minimum size=1.1em, draw=red, below= 1.5mm of B20]{};
\node[square, anchor=west,inner sep=2pt,inner sep=2.2pt,font=\ttfamily\small] (B21) at (B20.east) {1};
\node[circle,draw,fill=black!15,inner sep=1pt,minimum size=1.1em,below= 1.5mm of B21] {$d$};
\node[square, anchor=west,inner sep=2pt,inner sep=2.2pt,font=\ttfamily\small] (B22) at (B21.east) {0};
\draw[->,semithick] (B22.north) -- (S18.south);
\node[square, anchor=west,inner sep=2pt,inner sep=2.2pt,font=\ttfamily\small] (B23) at (B22.east) {0};
\node[square, anchor=west,inner sep=2pt,inner sep=2.2pt,font=\ttfamily\small] (B24) at (B23.east) {0};
\node[square, anchor=west,inner sep=2pt,inner sep=2.2pt,font=\ttfamily\small] (B25) at (B24.east) {0};
\node[square, anchor=west,inner sep=2pt,inner sep=2.2pt,font=\ttfamily\small] (B26) at (B25.east) {1};
\node[circle,draw,fill=black!15,inner sep=1pt,minimum size=1.1em,below= 1.5mm of B26] {$e$};
\node[square, anchor=west,inner sep=2pt,inner sep=2.2pt,font=\ttfamily\small] (B27) at (B26.east) {0};
\draw[->,semithick] (B27.north) -- (S22.south);
\node[square, anchor=west,inner sep=2pt,inner sep=2.2pt,font=\ttfamily\small] (B28) at (B27.east) {0};
\node[square, anchor=west,inner sep=2pt,inner sep=2.2pt,font=\ttfamily\small] (B29) at (B28.east) {0};
\node[square, anchor=west,inner sep=2pt,inner sep=2.2pt,font=\ttfamily\small] (B30) at (B29.east) {0};
\node[square, anchor=east,inner sep=2pt,inner sep=2.2pt,font=\ttfamily\small] (B12) at (B13.west) {1};
\node[circle,draw,fill=black!15,inner sep=1pt,minimum size=1.1em,below= 1.5mm of B12] {$b$};
\node[square, anchor=east,inner sep=2pt,inner sep=2.2pt,font=\ttfamily\small] (B11) at (B12.west) {0};
\node[square, anchor=east,inner sep=2pt,inner sep=2.2pt,font=\ttfamily\small] (B10) at (B11.west) {0};
\node[square, anchor=east,inner sep=2pt,inner sep=2.2pt,font=\ttfamily\small] (B9) at (B10.west) {0};
\node[square, anchor=east,inner sep=2pt,inner sep=2.2pt,font=\ttfamily\small] (B8) at (B9.west) {0};
\node[square, anchor=east,inner sep=2pt,inner sep=2.2pt,font=\ttfamily\small] (B7) at (B8.west) {0};
\node[square, anchor=east,inner sep=2pt,inner sep=2.2pt,font=\ttfamily\small] (B6) at (B7.west) {0};
\node[square, anchor=east,inner sep=2pt,inner sep=2.2pt,font=\ttfamily\small] (B5) at (B6.west) {0};
\node[square, anchor=east,inner sep=2pt,inner sep=2.2pt,font=\ttfamily\small] (B4) at (B5.west) {0};
\node[square, anchor=east,inner sep=2pt,inner sep=2.2pt,font=\ttfamily\small] (B3) at (B4.west) {0};
\node[square, anchor=east,inner sep=2pt,inner sep=2.2pt,font=\ttfamily\small] (B2) at (B3.west) {0};
\node[square, anchor=east,inner sep=2pt,inner sep=2.2pt,font=\ttfamily\small] (B1) at (B2.west) {0};
\draw[->,semithick] (B1.north) -- (S0.south);
\node[square, anchor=east,inner sep=2pt,inner sep=2.2pt,font=\ttfamily\small] (B0) at (B1.west) {1};
\node[circle,draw,fill=black!15,inner sep=1pt,minimum size=1.1em,below= 1.5mm of B0] {$a$};
\node[left= 0.3em of B0] {\normalsize $B$};
\end{tikzpicture}
    \caption{CAS representation of the temporal graph shown in Figure~\ref{fig:temporal-graphs}. CAS structure stores a sequence $S$ and a bitmap $B$. The $u$-th bit set to $1$ in $B$ marks the beginning of the $S$ block associated with the $u$-th source vertex.
Each block associated with the $u$-th source vertex is an event list ordered by time where, after a timestamp symbol $t$, there are target vertices symbols $v$ that, together, represent events of activation or
    deactivation of edges $(u, v)$ at timestamp $t$. The arrows illustrate the beginning of each $u$-th block in $S$ and the first corresponding $0$ in $B$.
    The vertex $c$ is red-crossed because no event leaves it.}\label{fig:cas}
\end{figure}
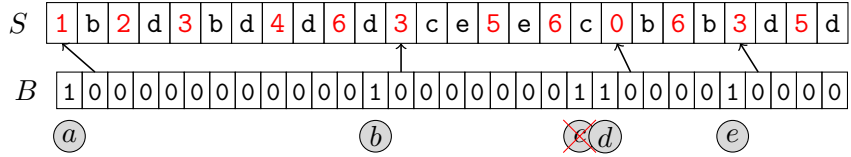

Compact Adjacency Sequence (CAS) also stores activation and deactivation events and uses the parity property to answer queries\footnote{Source code is available at \url{https://github.com/diegocaro/cas/}.}.
It determines the frequency of edge events in logarithmic time by using a wavelet tree~\citep{caro2015data}.

As illustrated in Figure~\ref{fig:cas}, CAS represents a temporal graph as a sequence $S = s_1, s_2, \ldots, s_n$ of symbols $s_i \in T \cup V$ and an extra bitmap $B$ of size $n + |V|$ for storage of information about adjacencies.

CAS partitions the sequence $S$ into blocks representing events $(t, v) \in T \times  V$ associated with each source vertex $u \in V$. The partitioning of $S$ follows two steps: (1) sort the events $(t, v)$ by the source vertices and by their timestamps, respectively; and (2) remove the timestamp repetitions.

The bitmap $B$ marks the starting position $k_i$ of the block associated with each  $i$-th source vertex in $S$ by filling with $k_i$ $0$'s after every $i$-th bit set to $1$.
In other words, the value of the first position of $B$ is $1$ and the values of the $k_1$ subsequent positions, associated with the first source vertex, are $0$.
The value of the next initialized position is $1$ and the values of the following $k_2$ positions, associated with the second source vertex, are $0$.
This pattern continues for each source vertex.
Note that CAS uses a sorted set $V$.
$B$ is implemented as a special abstract data type named bitvector, which provides two operations: $c=rank(B, s, i)$ and $j=select(B, s, c)$, enabling us to compute the count of a given symbol $s$ up to a position $i$ and to get the smallest position $j$ with count $c$ for symbol $s$ \citep{navarro2016compact}.
RRR is a well-known example\footnote{An implementation of RRR is found in \url{https://github.com/fclaude/libcds/blob/master/src/static/bitsequence/BitSequenceRRR.cpp}. Accessed in April, 1st 2020.} of succinct data structure to implement a bitvector with $rank()$ and $select()$ costing $O(1)$ operation \citep{raman2007succinct}.

In order to decrease space usage and improve query performance, CAS encodes the sequence $S$ using the wavelet tree~\citep{grossi2003high}, a succinct self-indexed structure that stores a sequence of $n$ symbols from an alphabet $\Sigma$ of size $\sigma$ using $O(n\log{\sigma})$ bits.
The standard version of wavelet trees uses a static $\Sigma$ that cannot change after its creation.

This data structure extends bitvector's $rank()$ and $select()$ to support the following queries with time complexity of $O(\log{\sigma})$: $rank_\alpha(S, i)$, $select_\alpha(S, f)$, $range\mathunderscore{}count_{[\alpha, \beta]}(S, i, j)$, $range\mathunderscore{}next\mathunderscore{}value_\alpha(S, i, j)$, and $range\mathunderscore{}next\mathunderscore{}value\mathunderscore{}pos_\alpha(S, i, j)$. Note that are alternative implementations in which using more space results in reducing time spent in some of these operations~\citep{navarro2016compact}.

The operation $rank_\alpha(S, i)$ retrieves the frequency of symbol $\alpha$ in the first $i$ elements of the sequence $S$;
$select_\alpha(S, f)$ returns the position $i$ such that the frequency of symbol $\alpha$ in $s_1, s_2, \ldots, s_i$ is $f$;
$range\mathunderscore{}count_{[\alpha, \beta]}(S, i, j)$ computes the number of symbols in $S= s_i, s_{i+1}, \ldots, s_{j}$ considering only symbols in the interval $[\alpha, \beta] \subseteq \Sigma$;
and $range\mathunderscore{}next\mathunderscore{}value_\alpha(S, i, j)$ gets the smallest symbol in $s_i, s_{i+1}, \ldots, s_{j}$ larger than $\alpha$.
Operation $range\mathunderscore{}next\mathunderscore{}value\mathunderscore{}pos_\alpha(S, i,j)$ returns the position of the symbol returned by $range\mathunderscore{}next\mathunderscore{}value_\alpha(S, i, j)$.

The wavelet tree also has the operation $range\mathunderscore{}report_{[\alpha, \beta]}(S, i, j)$, which retrieves separately the frequency of each distinct symbol in $S= s_i, s_{i+  1}, \ldots, s_{j}$ considering only symbols in $[\alpha, \beta] \subseteq \Sigma$~\citep{gagie2012new} with time complexity $O(k \log{\sigma})$, where $k$ is the number of symbols in the result.

CAS uses the wavelet tree by translating queries on the temporal graph represented by sequence $S$ into wavelet tree queries on bitmaps.
Rank and select queries on bitmaps have the same semantics of the equivalent queries on sequences and can be implemented with time complexity $O(1)$ and $o(\log n)$, respectively, where $n$ is the length of $B$, with little impact on space \citep{jacobson1988succinct}.
There are others ways to implement these queries where the tradeoff time-space complexity can be changed.

\subsubsection{Operation $\vect{has\mathunderscore{}edge}$ in CAS}

In CAS, an algorithm to answer \changed{\hasedgei} needs to find the frequency of $v$ in the block associated with $u$ in $S$ considering events in the interval $[t_{begin}, t_{end}]$.
First, this algorithm finds the beginning and ending positions, $i$ and $j$, of the block associated with the source vertex $u$ in $S$.

Given that there is a $1$ in $B$ for each vertex in $\mathcal{G}$, running operation $select_1(B, u)$ returns us the starting position of the $u$-th block in $B$. Then, when it runs $i = rank_0(B, select_1(B,u))$ the returned value $i$ points to the position of $u$ block in $S$.

Similarly, to get the ending position $j$ of the list of adjacencies of $u$ in $S$, the algorithm first executes operation $select_1(B, u+1)$ to get the position of the next symbol to $u$ in $B$. The number of symbols $0$ up to position $select_1(B, u+1)$ in $B$ is $rank_0(B, select_1(B, u+1))$, which returns a value pointing one position after the end of the event list of $u$ in $S$. Finally, the position $j$ is $rank_0(B, select_1(B, u+1)) -1$.

Next, to know the number of events before $t_{begin}$ and verify if $u$ is active when starts the target interval, the algorithm calls $k_{begin}=range\mathunderscore{}next\mathunderscore{}value\mathunderscore{}pos_{t_{begin}}(S, i, j)$ to find the first position $k_{begin}$ that has a timestamp symbol greater than $t_{begin}$ to restrict $S$ to the smaller block $S[i, k_{begin} - 1]$, which has only events that occurred until time $t_{begin}$.

Also, it calls $k_{end}=range\mathunderscore{}next\mathunderscore{}value\mathunderscore{}pos_{t_{end}}(S, i, j)$ to find the first position $k_{end}$ that has a timestamp symbol greater than $t_{end}$ to restrict $S$ to the block $S[k_{begin}, k_{end} - 1]$, which allows to count the number of events that occurred during the interval $[t_{begin}, t_{end}]$.

Finally, the algorithm calls $range\mathunderscore{}count_{[v, v]}(S, i, k_{begin} - 1)$ to count the frequency $f_{begin}$ of symbols $v$ in $S[i, k_{begin} -1]$ and $range\mathunderscore{}count_{[v, v]}(S, i, k_{end} - 1)$ to count the frequency $f_{end}$ of symbols $v$ in $S[k_{begin}, k_{end} -1]$.
If $f_{begin}$ is odd, then edge $(u, v)$ is active at time $t_{begin}$, else if $f_{end}$ is greater than zero, then $(u, v)$ is actived at some timestamp during the interval $[t_{begin}, t_{end}]$, otherwise $(u, v)$ is not active during $[t_{begin}, t_{end}]$.
Note that, if it is a point-based query, where $t = t_{begin} = t_{end}$ and $f = f_{begin} = f_{end}$, then the algorithm only needs to check if $f$ is odd to answer whether $(u, v)$ is active or not at timestamp $t$.

\subsubsection{Operation $\vect{neighbors}$ in CAS}
An algorithm to answer \changed{\neighsi} needs to get the frequency of all possible target vertices $v'$ in the block associated with $u$ in $S$ considering events from $t_{begin}$ to $t_{end}$.
Similar to \changed{\hasedgei}, it first finds the beginning and ending positions, $i$ and $j$, respectively, of the block associated with $u$.

Then, it finds the position $k_{begin}$ of the first symbol $t_{begin}$ by calling $k_{begin} =range\mathunderscore{}next\mathunderscore{}value\mathunderscore{}pos_{t_{begin}}(S, i, k_{end})$ and the position $k_{end}$ of the first symbol with value greater $t_{end}$ by calling $k_{end} = range\mathunderscore{}next\mathunderscore{}value\mathunderscore{}pos_{t_{end}}(S,
i, j)$ to restrict $S$ to the block $S[k_{begin}, k_{end} - 1]$.

Next, the algorithm calls $range\mathunderscore{}report_{[\alpha, \beta] \subseteq V}(S, 1, k_{begin})$ to collect into $C_1$ the frequency of all possible target vertices $v'$ inside $S[1, k_{begin}]$, and it also calls $range\mathunderscore{}report_{[\alpha, \beta] \subseteq V}(S, k_{begin} + 1, k_{end} - 1)$ to collect into $C_2$ the frequency of all possible target vertices inside $S[k_{begin} + 1, k_{end} - 1]$.

Finally, it checks weak or strong semantics for every vertex collected.
If the interval query has weak semantics, the algorithm only retrieves contacts in which edge $(u, v')$ enters active in $[t_{begin}, t_{end}]$ --- same as having odd frequency in $C_1$ ---  or whether it deactivates during $[t_{begin}, t_{end}]$ --- same as having frequency greater than $0$ in $C_2$.

Instead, if the interval query has strong semantics, a contact is returned if an edge $(u, v')$ enters active --- same as having odd frequency in $C_1$ --- and does not deactivate until $t_{end}$ --- same as having frequency equals to $0$ in $C_2$.

\subsubsection{Operation $\vect{neighbors^r}$ in CAS}

As the previous data structures, CAS also does not tackle \changed{\rneighsi} efficiently.
There are two different approaches to answer this query.
In the first approach, an algorithm checks the frequency of symbols $v$ in all blocks of $S$ associated with source vertices $u'$.
In order to determine these frequencies, first the algorithm restricts the symbols of each block, similarly to the previous algorithms. Next it calls $range\mathunderscore{}count_{[v, v] \subseteq V}$ inside each one, and, finally, it collects contacts in which an edge $(u', v)$ holds weak or strong semantics during $[t_{begin}, t_{end}]$.
A second approach, similar to previous structures, keeps another CAS structure where every edge would have its direction reversed.
Then, it answers \changed{\rneighsi} by calling \changed{\neighsi} in this second structure.
The first approach would access all structure, which would impact severely in time, while the second approach would double the space required.

 \subsection{Compact Events ordered by Time (CET)}
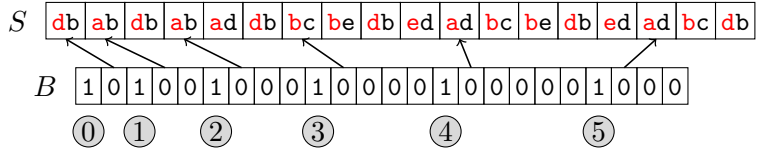
\begin{figure}
    \centering
\begin{tikzpicture}[square/.style={draw,outer sep=0pt,inner sep=2.2pt,text depth=.25ex,text height=1.75ex,font=\ttfamily\small}]
\node[square] (S0) {\textcolor{red}{d}b}
 node[square, right] (S2) at (S0.east) {\textcolor{red}{a}b}
 node[square, right] (S4) at (S2.east) {\textcolor{red}{d}b}
 node[square, right] (S6) at (S4.east) {\textcolor{red}{a}b}
 node[square, right] (S8) at (S6.east) {\textcolor{red}{a}d}
 node[square, right] (S10) at (S8.east) {\textcolor{red}{d}b}
 node[square, right] (S12) at (S10.east) {\textcolor{red}{b}c}
 node[square, right] (S14) at (S12.east) {\textcolor{red}{b}e}
 node[square, right] (S16) at (S14.east) {\textcolor{red}{d}b}
 node[square, right] (S18) at (S16.east) {\textcolor{red}{e}d}
 node[square, right] (S20) at (S18.east) {\textcolor{red}{a}d}
 node[square, right] (S22) at (S20.east) {\textcolor{red}{b}c}
 node[square, right] (S24) at (S22.east) {\textcolor{red}{b}e}
 node[square, right] (S26) at (S24.east) {\textcolor{red}{d}b}
 node[square, right] (S28) at (S26.east) {\textcolor{red}{e}d}
 node[square, right] (S30) at (S28.east) {\textcolor{red}{a}d}
 node[square, right] (S32) at (S30.east) {\textcolor{red}{b}c}
 node[square, right] (S34) at (S32.east) {\textcolor{red}{d}b};
\node[left= 0.3em of S0] {\normalsize $S$};
\node[square, anchor=north, yshift=-1em] (B13) at (S18.south) {0};
\node[square, anchor=west] (B14) at (B13.east) {1};
\node[circle,draw,fill=black!15,inner sep=1pt,below= 1.5mm of B14] {$4$};
\node[square, anchor=west] (B15) at (B14.east) {0};
\draw[->,semithick] (B15.north) -- (S20.south);
\node[square, anchor=west] (B16) at (B15.east) {0};
\node[square, anchor=west] (B17) at (B16.east) {0};
\node[square, anchor=west] (B18) at (B17.east) {0};
\node[square, anchor=west] (B19) at (B18.east) {0};
\node[square, anchor=west] (B20) at (B19.east) {1};
\node[circle,draw,fill=black!15,inner sep=1pt,below= 1.5mm of B20] {$5$};
\node[square, anchor=west] (B21) at (B20.east) {0};
\draw[->,semithick] (B21.north) -- (S30.south);
\node[square, anchor=west] (B22) at (B21.east) {0};
\node[square, anchor=west] (B23) at (B22.east) {0};
\node[square, anchor=east] (B12) at (B13.west) {0};
\node[square, anchor=east] (B11) at (B12.west) {0};
\node[square, anchor=east] (B10) at (B11.west) {0};
\draw[->,semithick] (B10.north) -- (S12.south);
\node[square, anchor=east] (B9) at (B10.west) {1};
\node[circle,draw,fill=black!15,inner sep=1pt,below= 1.5mm of B9] {$3$};
\node[square, anchor=east] (B8) at (B9.west) {0};
\node[square, anchor=east] (B7) at (B8.west) {0};
\node[square, anchor=east] (B6) at (B7.west) {0};
\draw[->,semithick] (B6.north) -- (S6.south);
\node[square, anchor=east] (B5) at (B6.west) {1};
\node[circle,draw,fill=black!15,inner sep=1pt,below= 1.5mm of B5] {$2$};
\node[square, anchor=east] (B4) at (B5.west) {0};
\node[square, anchor=east] (B3) at (B4.west) {0};
\draw[->,semithick] (B3.north) -- (S2.south);
\node[square, anchor=east] (B2) at (B3.west) {1};
\node[circle,draw,fill=black!15,inner sep=1pt,below= 1.5mm of B2] {$1$};
\node[square, anchor=east] (B1) at (B2.west) {0};
\draw[->,semithick] (B1.north) -- (S0.south);
\node[square, anchor=east] (B0) at (B1.west) {1};
\node[circle,draw,fill=black!15,inner sep=1pt,below= 1.5mm of B0] {$0$};
;
\node[left= 0.3em of B0] {\normalsize $B$};
\end{tikzpicture}
    \caption{CET representation of the temporal graph shown in Figure~\ref{fig:temporal-graphs}. CET structure has a sequence $S^{(2)}$ containing $2$-dimensional symbols and a bitmap $B$. The $t$-th bit set to $1$ in $B$ marks the beginning of the $S^{(2)}$ block associated with the $t$-th timestamp. Each block associated with the $t$-th timestamp has symbols $(u, v) \in V\times V$ that represents an event of activation or deactivation of edge $(u, v)$ at timestamp $t$. The arrows illustrate the beginning of each $t$-th block in $S^{(2)}$ and the first corresponding $0$ in $B$.}\label{fig:cet}
\end{figure}

Compact Events ordered by Time (CET) uses a $2$-dimensional sequence $S^{(2)} = s^{(2)}_1, s^{(2)}_2, \ldots$, where symbols $s^{(2)}_i$ represent tuples $(u, v) \in V\times V$, and a bitmap $B$ with size $|B| = |S^{(2)}| + |T|$ to mark timestamps~\citep{caro2015data}\footnote{Source code is available at \url{https://github.com/diegocaro/cet/}.}.
As illustrated in Figure~\ref{fig:cet}, it groups symbols in $S^{(2)}$ by time instead of grouping by source vertex.
Each symbol $s^{(2)}_i = (u, v)$ in a block associated with time $t$ represents an event of activation or deactivation of edge $(u, v)$ at time $t$.
Bitmap $B$ marks the beginning of events associated with time $t$.

CET uses the interleaved wavelet tree data structure to store the sequence $S^{(2)}$ efficiently~\citep{caro2015data}.
This structure generalizes the operations supported by standard wavelet trees to multidimensional sequences $S^{(d)} = s^{(d)}_1, s^{(d)}_2, \ldots$, where $d$ is the dimensionality of symbols $s^{(d)} = (s_1, s_2, \ldots, s_d)$.

CET uses this structure to store edges $(u, v)$ efficiently as $2$-dimensional sequences and to retrieve the frequency of edges by using queries such as $rank_{(u, v)}(S^{(2)}, i)$, $range\mathunderscore{}count_{\Sigma^{(2)}}(S^{(2)}, i, j)$, $range\mathunderscore{}report_{\Sigma^{(2)}}(S^{(2)}, i, j)$,
$range\mathunderscore{}next\mathunderscore{}value_{(u, v)}(S^{(2)}, i, j)$, and $range\mathunderscore{}next\mathunderscore{}value\mathunderscore{}pos_{(u, v)}(S^{(2)}, i, j)$ in logarithmic time $O(\log{\Sigma^{(2)}})$, where $\Sigma^{(2)}$ is the alphabet formed by every pair of vertices $V\times V$.

\subsubsection{Operation $\vect{has\mathunderscore{}edge}$ in CET}

An algorithm to answer \changed{\hasedgei} first finds the positions $i$ and $j$ in $S^{(2)}$ associated with $t_{begin}+1$ and $t_{end} + 1$, respectively, by using the operation $select()$ of the bitvector implementing the bitmap $B$. 

Then, it calls $rank_{(u,v)}(S^{(2)}, i - 1)$ to retrieve the frequency of events regarding edge $(u, v)$ until timestamp $t_{begin}$ and $rank_{(u,v)}(S^{(2)}, j - 1)$ to retrieve the frequency of events during the $[t_{begin}, t_{end}]$.
Next, similar to previous strategies, it uses the parity property to answer whether edge $(u, v)$ is active or not during the interval $[t_{begin}, t_{end}]$.

\subsubsection{Operation $\vect{neighbors}$ in CET}

In order to answer \changed{\neighsi}, an algorithm finds positions $i$ and $j$ associated to $t$ and $t + 1$, then it performs $range\mathunderscore{}report_{(u, v) \subseteq V}(S^{(2)}, 1, i-1)$.
For weak semantics, if the frequency of symbol $(u,v)$ is odd, it adds edge $(u, v)$ to the result, then, for the remaining symbols, it calls $range\mathunderscore{}report_{(u, v') \subseteq V}(S^{(2)}, i, j-1)$ and, if the frequency of symbol $(u',v')$ is greater than $0$, it also adds edge $(u', v')$ to the result.
For strong semantics, if the frequency of symbol $(u,v)$ is odd, it discards edge $(u, v)$ from the result, then, for the remaining symbols, it calls $range\mathunderscore{}report_{(u, v) \subseteq V}(S^{(2)}, i, j-1)$ and removes from result the edges in which the corresponding symbols have frequency equal to $0$.


\subsubsection{Operation $\vect{neighbors^r}$ in CET}

Differently from the previous structures, CET has the same time complexity for retrieving direct and reverse neighbors of a given vertex $v$.
An algorithm to answer \changed{\rneighsi} is similar to $neighbors(\mathcal{G}, v, t_{begin}, t_{end})$.
The only difference is that it calls $range\mathunderscore{}report_{(u, v) \subseteq V}$ instead of $range\mathunderscore{}report_{(v, u) \subseteq V}$.
In other words, we just swap the dimension values of symbols in $S^{(2)}$.


\subsection{Temporal Graph Compressed Suffix Array}

Temporal Graph Compressed Suffix Array (TGCSA) is a technique based on the Compressed Suffix Array (CSA)~\citep{csa} to store and query temporal graphs~\citep{brisaboa2018using}\footnote{Source code of CSA is available at \url{https://github.com/diegocaro/sdsl-lite/}.}.
It represents a list of contacts using a string with unique characteristics and transforms the problem of querying temporal graphs in a substring matching problem.
Therefore, TGCSA represents a list of contacts $C = c_1, c_2, \ldots, c_n$, where $c_i = \{u, v, t_{begin}, t_{end}\} \in \mathcal{G}$, as a sequence $S = s_1, s_2, \ldots, s_m$ formed by the concatenation of all $c_i$.

Note that, each element in a contact should be represented by an unique symbol.
For this, TGCSA constructs and stores a dictionary $\Sigma$ to encode $S$ considering the following rules: $u \in [1, |V|]$, $v \in [|V| + 1, 2 |V|]$, $t_{begin} \in [2|V| + 1, 2|V| + |T|]$ and $t_{end} \in [2|V| + |T| + 1, 2|V| + 2|T|]$.
That is, symbols that encode source vertex have lower values than symbols that encode target vertex, which in turn, have lower values than activation timestamps and, which in turn, have lower values than deactivation timestamps.
By using these rules, TGCSA can order the symbols in $\Sigma$ in $4$ different groups and take advantage of this property for speeding-up queries later.
For now on, we will assume the string $E$ to be the sequence of encoded contacts obtained from $S$ by using codes in $\Sigma$.
Note that, it is possible to decode symbols in $E$ using $\Sigma$.

\begin{figure}
\begin{adjustwidth}{-2in}{-2in}
  \centering
\begin{tikzpicture}[square/.style={draw,outer sep=0pt,inner sep=1.85pt,text depth=.25ex,text height=1.75ex,font=\scriptsize\ttfamily,minimum width=4.3mm}]
\begin{scope}[opacity=0.5]
\node[square] (S0) {a}
 node[square, right] (S1) at (S0.east) {b}
 node[square, right] (S2) at (S1.east) {1}
 node[square, right] (S3) at (S2.east) {3}
 node[square, right] (S4) at (S3.east) {a}
 node[square, right] (S5) at (S4.east) {d}
 node[square, right] (S6) at (S5.east) {2}
 node[square, right] (S7) at (S6.east) {3}
 node[square, right] (S8) at (S7.east) {a}
 node[square, right] (S9) at (S8.east) {d}
 node[square, right] (S10) at (S9.east) {4}
 node[square, right] (S11) at (S10.east) {6}
 node[square, right] (S12) at (S11.east) {b}
 node[square, right] (S13) at (S12.east) {c}
 node[square, right] (S14) at (S13.east) {3}
 node[square, right] (S15) at (S14.east) {6}
 node[square, right] (S16) at (S15.east) {b}
 node[square, right] (S17) at (S16.east) {e}
 node[square, right] (S18) at (S17.east) {3}
 node[square, right] (S19) at (S18.east) {5}
 node[square, right] (S20) at (S19.east) {d}
 node[square, right] (S21) at (S20.east) {b}
 node[square, right] (S22) at (S21.east) {0}
 node[square, right] (S23) at (S22.east) {6}
 node[square, right] (S24) at (S23.east) {e}
 node[square, right] (S25) at (S24.east) {d}
 node[square, right] (S26) at (S25.east) {3}
 node[square, right] (S27) at (S26.east) {5};
\node[left= 0.3em of S0] {\normalsize $S$};
\node[square, anchor=north, yshift=-3mm] (E0) at (S0.south) {0}
 node[square, right] (E1) at (E0.east) {4}
 node[square, right] (E2) at (E1.east) {9}
 node[square, right] (E3) at (E2.east) {13}
 node[square, right] (E4) at (E3.east) {0}
 node[square, right] (E5) at (E4.east) {6}
 node[square, right] (E6) at (E5.east) {10}
 node[square, right] (E7) at (E6.east) {13}
 node[square, right] (E8) at (E7.east) {0}
 node[square, right] (E9) at (E8.east) {6}
 node[square, right] (E10) at (E9.east) {12}
 node[square, right] (E11) at (E10.east) {15}
 node[square, right] (E12) at (E11.east) {1}
 node[square, right] (E13) at (E12.east) {5}
 node[square, right] (E14) at (E13.east) {11}
 node[square, right] (E15) at (E14.east) {15}
 node[square, right] (E16) at (E15.east) {1}
 node[square, right] (E17) at (E16.east) {7}
 node[square, right] (E18) at (E17.east) {11}
 node[square, right] (E19) at (E18.east) {14}
 node[square, right] (E20) at (E19.east) {2}
 node[square, right] (E21) at (E20.east) {4}
 node[square, right] (E22) at (E21.east) {8}
 node[square, right] (E23) at (E22.east) {15}
 node[square, right] (E24) at (E23.east) {3}
 node[square, right] (E25) at (E24.east) {6}
 node[square, right] (E26) at (E25.east) {11}
 node[square, right] (E27) at (E26.east) {14};
\node[left= 0.3em of E0] {\normalsize $E$};
\node[black!90] (EL0) at ($(E0) - (0, 0.3)$) {\tiny 0};
\node[black!90] (EL1) at ($(E1) - (0, 0.3)$) {\tiny 1};
\node[black!90] (EL2) at ($(E2) - (0, 0.3)$) {\tiny 2};
\node[black!90] (EL3) at ($(E3) - (0, 0.3)$) {\tiny 3};
\node[black!90] (EL4) at ($(E4) - (0, 0.3)$) {\tiny 4};
\node[black!90] (EL5) at ($(E5) - (0, 0.3)$) {\tiny 5};
\node[black!90] (EL6) at ($(E6) - (0, 0.3)$) {\tiny 6};
\node[black!90] (EL7) at ($(E7) - (0, 0.3)$) {\tiny 7};
\node[black!90] (EL8) at ($(E8) - (0, 0.3)$) {\tiny 8};
\node[black!90] (EL9) at ($(E9) - (0, 0.3)$) {\tiny 9};
\node[black!90] (EL10) at ($(E10) - (0, 0.3)$) {\tiny 10};
\node[black!90] (EL11) at ($(E11) - (0, 0.3)$) {\tiny 11};
\node[black!90] (EL12) at ($(E12) - (0, 0.3)$) {\tiny 12};
\node[black!90] (EL13) at ($(E13) - (0, 0.3)$) {\tiny 13};
\node[black!90] (EL14) at ($(E14) - (0, 0.3)$) {\tiny 14};
\node[black!90] (EL15) at ($(E15) - (0, 0.3)$) {\tiny 15};
\node[black!90] (EL16) at ($(E16) - (0, 0.3)$) {\tiny 16};
\node[black!90] (EL17) at ($(E17) - (0, 0.3)$) {\tiny 17};
\node[black!90] (EL18) at ($(E18) - (0, 0.3)$) {\tiny 18};
\node[black!90] (EL19) at ($(E19) - (0, 0.3)$) {\tiny 19};
\node[black!90] (EL20) at ($(E20) - (0, 0.3)$) {\tiny 20};
\node[black!90] (EL21) at ($(E21) - (0, 0.3)$) {\tiny 21};
\node[black!90] (EL22) at ($(E22) - (0, 0.3)$) {\tiny 22};
\node[black!90] (EL23) at ($(E23) - (0, 0.3)$) {\tiny 23};
\node[black!90] (EL24) at ($(E24) - (0, 0.3)$) {\tiny 24};
\node[black!90] (EL25) at ($(E25) - (0, 0.3)$) {\tiny 25};
\node[black!90] (EL26) at ($(E26) - (0, 0.3)$) {\tiny 26};
\node[black!90] (EL27) at ($(E27) - (0, 0.3)$) {\tiny 27};
\node[red] at ($(S2.north west) + (0, 0.2)$) {$c_{0}$};
\draw[red, thick] ($(S4.north west) + (0, 0.2)$) -- ($(E4.south west) - (0, 0.2)$);
\node[red] at ($(S6.north west) + (0, 0.2)$) {$c_{1}$};
\draw[red, thick] ($(S8.north west) + (0, 0.2)$) -- ($(E8.south west) - (0, 0.2)$);
\node[red] at ($(S10.north west) + (0, 0.2)$) {$c_{2}$};
\draw[red, thick] ($(S12.north west) + (0, 0.2)$) -- ($(E12.south west) - (0, 0.2)$);
\node[red] at ($(S14.north west) + (0, 0.2)$) {$c_{3}$};
\draw[red, thick] ($(S16.north west) + (0, 0.2)$) -- ($(E16.south west) - (0, 0.2)$);
\node[red] at ($(S18.north west) + (0, 0.2)$) {$c_{4}$};
\draw[red, thick] ($(S20.north west) + (0, 0.2)$) -- ($(E20.south west) - (0, 0.2)$);
\node[red] at ($(S22.north west) + (0, 0.2)$) {$c_{5}$};
\draw[red, thick] ($(S24.north west) + (0, 0.2)$) -- ($(E24.south west) - (0, 0.2)$);
\node[red] at ($(S26.north west) + (0, 0.2)$) {$c_{6}$};
\node[square, anchor=north, yshift=-10mm] (A0) at (E0.south) {0}
 node[square, right] (A1) at (A0.east) {4}
 node[square, right] (A2) at (A1.east) {8}
 node[square, right] (A3) at (A2.east) {12}
 node[square, right] (A4) at (A3.east) {16}
 node[square, right] (A5) at (A4.east) {20}
 node[square, right] (A6) at (A5.east) {24}
 node[square, right] (A7) at (A6.east) {21}
 node[square, right] (A8) at (A7.east) {1}
 node[square, right] (A9) at (A8.east) {13}
 node[square, right] (A10) at (A9.east) {5}
 node[square, right] (A11) at (A10.east) {25}
 node[square, right] (A12) at (A11.east) {9}
 node[square, right] (A13) at (A12.east) {17}
 node[square, right] (A14) at (A13.east) {22}
 node[square, right] (A15) at (A14.east) {2}
 node[square, right] (A16) at (A15.east) {6}
 node[square, right] (A17) at (A16.east) {26}
 node[square, right] (A18) at (A17.east) {18}
 node[square, right] (A19) at (A18.east) {14}
 node[square, right] (A20) at (A19.east) {10}
 node[square, right] (A21) at (A20.east) {3}
 node[square, right] (A22) at (A21.east) {7}
 node[square, right] (A23) at (A22.east) {19}
 node[square, right] (A24) at (A23.east) {27}
 node[square, right] (A25) at (A24.east) {11}
 node[square, right] (A26) at (A25.east) {15}
 node[square, right] (A27) at (A26.east) {23};
\node[left= 0.3em of A0] {\normalsize $A$};
\draw (A0.north) edge [->, line width=0.5pt, out=90, in=270,looseness=0.15] (EL0.south);
\draw (A1.north) edge [->, line width=0.5pt, out=90, in=270,looseness=0.15] (EL4.south);
\draw (A2.north) edge [->, line width=0.5pt, out=90, in=270,looseness=0.15] (EL8.south);
\draw (A3.north) edge [->, line width=0.5pt, out=90, in=270,looseness=0.15] (EL12.south);
\draw (A4.north) edge [->, line width=0.5pt, out=90, in=270,looseness=0.15] (EL16.south);
\draw (A5.north) edge [->, line width=0.5pt, out=90, in=270,looseness=0.15] (EL20.south);
\draw (A6.north) edge [->, line width=0.5pt, out=90, in=270,looseness=0.15] (EL24.south);
\draw (A7.north) edge [->, line width=0.5pt, red, out=90, in=270,looseness=0.3] (EL21.south);
\draw (A8.north) edge [->, line width=0.5pt, red, out=90, in=270,looseness=0.3] (EL1.south);
\end{scope}
\node[square, anchor=north, yshift=-3mm] (D0) at (A0.south) {1}
 node[square, right] (D1) at (D0.east) {0}
 node[square, right] (D2) at (D1.east) {0}
 node[square, right] (D3) at (D2.east) {1}
 node[square, right] (D4) at (D3.east) {0}
 node[square, right] (D5) at (D4.east) {1}
 node[square, right] (D6) at (D5.east) {1}
 node[square, right] (D7) at (D6.east) {1}
 node[square, right] (D8) at (D7.east) {0}
 node[square, right] (D9) at (D8.east) {1}
 node[square, right] (D10) at (D9.east) {1}
 node[square, right] (D11) at (D10.east) {0}
 node[square, right] (D12) at (D11.east) {0}
 node[square, right] (D13) at (D12.east) {1}
 node[square, right] (D14) at (D13.east) {1}
 node[square, right] (D15) at (D14.east) {1}
 node[square, right] (D16) at (D15.east) {1}
 node[square, right] (D17) at (D16.east) {1}
 node[square, right] (D18) at (D17.east) {0}
 node[square, right] (D19) at (D18.east) {0}
 node[square, right] (D20) at (D19.east) {1}
 node[square, right] (D21) at (D20.east) {1}
 node[square, right] (D22) at (D21.east) {0}
 node[square, right] (D23) at (D22.east) {1}
 node[square, right] (D24) at (D23.east) {0}
 node[square, right] (D25) at (D24.east) {1}
 node[square, right] (D26) at (D25.east) {0}
 node[square, right] (D27) at (D26.east) {0};
\node[left= 0.3em of D0] {\normalsize $D$};
\node[black!90] (DL0) at ($(D0) - (0, 0.3)$) {\tiny a};
\node[black!90] (DL1) at ($(D1) - (0, 0.3)$) {\tiny a};
\node[black!90] (DL2) at ($(D2) - (0, 0.3)$) {\tiny a};
\node[black!90] (DL3) at ($(D3) - (0, 0.3)$) {\tiny b};
\node[black!90] (DL4) at ($(D4) - (0, 0.3)$) {\tiny b};
\node[black!90] (DL5) at ($(D5) - (0, 0.3)$) {\tiny d};
\node[black!90] (DL6) at ($(D6) - (0, 0.3)$) {\tiny e};
\node[black!90] (DL7) at ($(D7) - (0, 0.3)$) {\tiny b};
\node[black!90] (DL8) at ($(D8) - (0, 0.3)$) {\tiny b};
\node[black!90] (DL9) at ($(D9) - (0, 0.3)$) {\tiny c};
\node[black!90] (DL10) at ($(D10) - (0, 0.3)$) {\tiny d};
\node[black!90] (DL11) at ($(D11) - (0, 0.3)$) {\tiny d};
\node[black!90] (DL12) at ($(D12) - (0, 0.3)$) {\tiny d};
\node[black!90] (DL13) at ($(D13) - (0, 0.3)$) {\tiny e};
\node[black!90] (DL14) at ($(D14) - (0, 0.3)$) {\tiny 0};
\node[black!90] (DL15) at ($(D15) - (0, 0.3)$) {\tiny 1};
\node[black!90] (DL16) at ($(D16) - (0, 0.3)$) {\tiny 2};
\node[black!90] (DL17) at ($(D17) - (0, 0.3)$) {\tiny 3};
\node[black!90] (DL18) at ($(D18) - (0, 0.3)$) {\tiny 3};
\node[black!90] (DL19) at ($(D19) - (0, 0.3)$) {\tiny 3};
\node[black!90] (DL20) at ($(D20) - (0, 0.3)$) {\tiny 4};
\node[black!90] (DL21) at ($(D21) - (0, 0.3)$) {\tiny 3};
\node[black!90] (DL22) at ($(D22) - (0, 0.3)$) {\tiny 3};
\node[black!90] (DL23) at ($(D23) - (0, 0.3)$) {\tiny 5};
\node[black!90] (DL24) at ($(D24) - (0, 0.3)$) {\tiny 5};
\node[black!90] (DL25) at ($(D25) - (0, 0.3)$) {\tiny 6};
\node[black!90] (DL26) at ($(D26) - (0, 0.3)$) {\tiny 6};
\node[black!90] (DL27) at ($(D27) - (0, 0.3)$) {\tiny 6};
\node[square, anchor=north, yshift=-3mm] (P0) at (D0.south) {8}
 node[square, right] (P1) at (P0.east) {10}
 node[square, right] (P2) at (P1.east) {12}
 node[square, right] (P3) at (P2.east) {9}
 node[square, right] (P4) at (P3.east) {13}
 node[square, right] (P5) at (P4.east) {7}
 node[square, right] (P6) at (P5.east) {11}
 node[square, right] (P7) at (P6.east) {14}
 node[square, right] (P8) at (P7.east) {15}
 node[square, right] (P9) at (P8.east) {19}
 node[square, right] (P10) at (P9.east) {16}
 node[square, right] (P11) at (P10.east) {17}
 node[square, right] (P12) at (P11.east) {20}
 node[square, right] (P13) at (P12.east) {18}
 node[square, right] (P14) at (P13.east) {27}
 node[square, right] (P15) at (P14.east) {21}
 node[square, right] (P16) at (P15.east) {22}
 node[square, right] (P17) at (P16.east) {24}
 node[square, right] (P18) at (P17.east) {23}
 node[square, right] (P19) at (P18.east) {26}
 node[square, right] (P20) at (P19.east) {25}
 node[square, right] (P21) at (P20.east) {1}
 node[square, right] (P22) at (P21.east) {2}
 node[square, right] (P23) at (P22.east) {5}
 node[square, right] (P24) at (P23.east) {28}
 node[square, right] (P25) at (P24.east) {3}
 node[square, right] (P26) at (P25.east) {4}
 node[square, right] (P27) at (P26.east) {6};
\node[left= 0.3em of P0] {\normalsize $\Psi$};
\node[black!90] (PL0) at ($(P0) - (0, 0.3)$) {\tiny 0};
\node[black!90] (PL1) at ($(P1) - (0, 0.3)$) {\tiny 1};
\node[black!90] (PL2) at ($(P2) - (0, 0.3)$) {\tiny 2};
\node[black!90] (PL3) at ($(P3) - (0, 0.3)$) {\tiny 3};
\node[black!90] (PL4) at ($(P4) - (0, 0.3)$) {\tiny 4};
\node[black!90] (PL5) at ($(P5) - (0, 0.3)$) {\tiny 5};
\node[black!90] (PL6) at ($(P6) - (0, 0.3)$) {\tiny 6};
\node[black!90] (PL7) at ($(P7) - (0, 0.3)$) {\tiny 7};
\node[black!90] (PL8) at ($(P8) - (0, 0.3)$) {\tiny 8};
\node[black!90] (PL9) at ($(P9) - (0, 0.3)$) {\tiny 9};
\node[black!90] (PL10) at ($(P10) - (0, 0.3)$) {\tiny 10};
\node[black!90] (PL11) at ($(P11) - (0, 0.3)$) {\tiny 11};
\node[black!90] (PL12) at ($(P12) - (0, 0.3)$) {\tiny 12};
\node[black!90] (PL13) at ($(P13) - (0, 0.3)$) {\tiny 13};
\node[black!90] (PL14) at ($(P14) - (0, 0.3)$) {\tiny 14};
\node[black!90] (PL15) at ($(P15) - (0, 0.3)$) {\tiny 15};
\node[black!90] (PL16) at ($(P16) - (0, 0.3)$) {\tiny 16};
\node[black!90] (PL17) at ($(P17) - (0, 0.3)$) {\tiny 17};
\node[black!90] (PL18) at ($(P18) - (0, 0.3)$) {\tiny 18};
\node[black!90] (PL19) at ($(P19) - (0, 0.3)$) {\tiny 19};
\node[black!90] (PL20) at ($(P20) - (0, 0.3)$) {\tiny 20};
\node[black!90] (PL21) at ($(P21) - (0, 0.3)$) {\tiny 21};
\node[black!90] (PL22) at ($(P22) - (0, 0.3)$) {\tiny 22};
\node[black!90] (PL23) at ($(P23) - (0, 0.3)$) {\tiny 23};
\node[black!90] (PL24) at ($(P24) - (0, 0.3)$) {\tiny 24};
\node[black!90] (PL25) at ($(P25) - (0, 0.3)$) {\tiny 25};
\node[black!90] (PL26) at ($(P26) - (0, 0.3)$) {\tiny 26};
\node[black!90] (PL27) at ($(P27) - (0, 0.3)$) {\tiny 27};
\draw (P0.north) edge [->, line width=0.5pt, bend left] (A0.south);
\draw (P8.north) edge [->, line width=0.5pt, bend left] (A8.south);
\draw (P15.north) edge [->, line width=0.5pt, bend left] (A15.south);
\draw (P21.north) edge [->, line width=0.5pt, bend left] (A21.south);
\draw (PL0.south) edge [->, line width=0.5pt, out=270, in=270, looseness=0.2] ($(PL8.south) - (0.08, 0)$);
\draw (PL8.south) edge [->, line width=0.5pt, out=270, in=270, looseness=0.2] ($(PL15.south) - (0.08, 0)$);
\draw (PL15.south) edge [->, line width=0.5pt, out=270, in=270, looseness=0.2] ($(PL21.south) - (0.08, 0)$);
\draw (PL21.south) edge [->, line width=0.5pt, out=270, in=270, looseness=0.2] ($(PL1.south) + (0.08, 0)$);
\node [anchor=south, yshift=5mm, xshift=-1.5mm] (DIC) at (S14.north) {
\scriptsize\ttfamily
\setlength{\tabcolsep}{0.4em}%
\def\arraystretch{0.55}%
\begin{tabular}{|cc|cc|cc|cc|} \hline
\multicolumn{2}{|c}{$u$} & \multicolumn{2}{c}{$v$} & \multicolumn{2}{c}{$t_{begin}$} & \multicolumn{2}{c|}{$t_{end}$} \\ \hline
a & 0 & b & 4 & 0 & 8 & 4 & 12 \\ \hline
b & 1 & c & 5 & 1 & 9 & 3 & 13 \\ \hline
d & 2 & d & 6 & 2 & 10 & 5 & 14 \\ \hline
e & 3 & e & 7 & 3 & 11 & 6 & 15 \\ \hline
\end{tabular}
};
\node[left= 0em of DIC] {\normalsize $\Sigma$};
\end{tikzpicture}
\end{adjustwidth}
    \caption{TGCSA representation of the temporal graph shown in Figure~\ref{fig:temporal-graphs}. Temporal graph representation based on the uncompressed suffix array is represented by the structures with light gray color: the sequences $S$ and $E$, and the suffix array $A$. Sequence $S$ is formed by concatenating all contacts of the temporal graph. Sequence $E$ has the encodings of symbols in $S$ using the dictionary $\Sigma$, and $A$ is obtained by ordering lexicographically the suffixes in $E$. TGCSA stores only some structures: a dictionary $\Sigma$, a bit array $D$, and a successor suffix array $\Psi$. $D$ keeps track of $A$ symbols by setting $1$ at positions that symbols first appear in $A$, $\Psi$ preserves the property $A[\Psi[i]] = A[i] + 1$, thus it is possible to retrieve the next symbol in the original sequence, and $\Sigma$ is used to map original and encoded symbols. The arrows represent association examples between two different arrays or, in the case of $\Psi$, the application of sucessively calling itself to find the next symbols of a contact.}\label{fig:csa}
\end{figure}
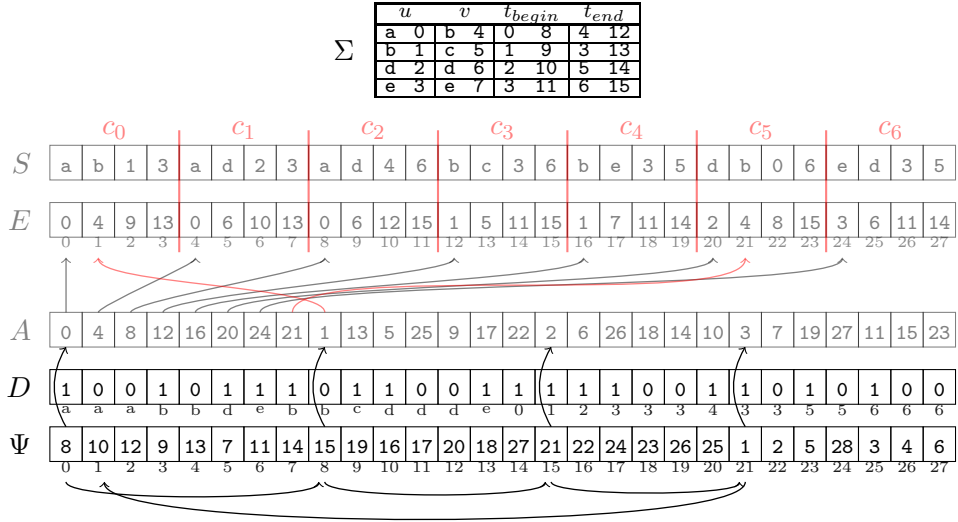

The standard Sufix Array (SA) strategy\citep{manber1993suffix}, depicted in light gray in Figure~\ref{fig:csa},  enumerates the suffixes from the string $E$ and sorts them into an array of integers $A$.
This approach allows finding substrings in $E$ that match a given query in logarithmic time on the length of $E$ 
by first encoding the substring query using $\Sigma$ and, then, performing binary search in $A$ to retrieve the possible matches.
Moreover, this strategy can also retrieve the context of a match in constant time by accessing the surrounding symbols of $E$.

However, $E$ and $A$ can consume much space and, thus, in cases in which space is important, the compressed suffix array (CSA) strategy is more appropriate.
As show in darker color in Figure~\ref{fig:csa}, instead of storing $E$ and $A$, CSA stores a bitvector $D$ in the format $10^{f_1}10^{f_2}\ldots10^{f_{|\Sigma|}}$, where $f_i$ is the frequency of the $i$-th symbol in $\Sigma$, and the successor array $\Psi$ with the property $A[\Psi[i]] = A[i] + 1$ to keep track of the next suffixes and, thus, the next positions of $E$ symbols.
Array $\Psi$ can be further compressed by applying DeltaGap to each \textit{run}, i.e each already sorted part.

In order to find matches using the CSA data structure given a query substring $Q = q_1, q_2, \ldots, q_l$, an algorithm first encodes $Q$ into the substring $E' = e'_1, e'_2, \ldots, e'_l$ using $\Sigma$, 
then it calls $i = select_1(D, e'_1)$ and $j = select_1(D, e'_1 + 1) - 1$ to find the range of positions $[i, j]$ containing suffix candidates beginning with the symbol $e'_1$.
Next, for each suffix candidate $Su$ with starting position at $k \in [i, j]$, it tries to retrieve its next symbol position by calling $k = \Psi[k]$ and, then, its next symbol value by calling $rank_1(D, k - 1)$.
Next symbols for a candidate $Su$ are successively retrieved
until the algorithm finds a symbol that does not match the next symbol in the query $E'$ or it successfully matches its first $l$ symbols, \textit{i.e.} the query size.
In the first case, it discards safely the corresponding suffix candidate and, in the second, it returns the corresponding matching suffix $Su$.


\subsubsection{Operation $\vect{has\mathunderscore{}edge}$ in TGCSA}

An algorithm to answer \changed{\hasedgei} is similar to the string matching strategy we described.
First, it encodes $u$ to its corresponding encoding $e'$ using the dictionary $\Sigma$.
Then, it calls $i = select_1(D, e')$ and $j = select_1(D, e' + 1) - 1$ to find the beginning and ending positions of suffixes that start with the encoded symbol $e'$.
Next, for each suffix candidate $Su$ with stating position at $k = [i, j]$, it tries to construct a candidate contact $c = \{ u', v', t_{begin}', t_{end}' \}$, where $u' = u$, $v' = M(rank_1(D, \Psi[k]) - 1)$, $t_{begin}' = M(rank_1(D, \Psi[\Psi[k]]) - 1)$ and $t_{end}' = M(rank_1(D, \Psi[\Psi[\Psi[k]]]) - 1)$, where the function $M(.)$ decodes symbols using $\Sigma$.
Note that, $k = \Psi[k]$ retrieves the position of the next encoded symbol, $rank_1(D, k) - 1$ computes the encoded symbol at position $k$ and $M(.)$ decodes it back to the corresponding contact element.
Finally, it collects all candidate contacts that intervals overlaps with $[t_{begin}, t_{end}]$.
As the general substring matching case we presented earlier, the algorithm can also discard candidates that do not match the query immediately as next symbols are discovered.

\subsubsection{Operation $\vect{neighbors}$ in TGCSA}

An algorithm to answer \changed{\neighsi}, similarly, encodes $u$ to its corresponding encoding $e'$ using the dictionary $\Sigma$, finds the sufixes that start with the encoded symbol $e'$ and constructs the candidate contacts.
However, it collects only the candidate contacts that satisfy the required interval semantics.
If $neighbors(\mathcal{G}, u, t_{begin}, t_{end})$ has weak semantics, then the algorithm collects the contacts in which $t_{begin}' \leq t_{end}$ and $t_{end}' \geq t_{begin}$, otherwise, if it has strong semantics, then it collects the contacts in which $t_{begin}' \leq t_{begin} \leq t_{end} \leq t_{end}'$.

\subsubsection{Operation $\vect{neighbors^r}$ in TGCSA}

In order to answer \changed{\rneighsi} efficiently, the authors introduced a modified version of the $\Psi$ array.
In the original array $\Psi$, if a suffix candidate $Su$ with starting position at $k$ starts with a symbol that corresponds to a deactivation timestamp encoding, then, the next suffix symbol at position $k = \Psi[k]$ would correspond to the source vertex encoding of the next contact.
In \citep{brisaboa2018using}, the authors modified $\Psi$ to make it cyclical regarding the same contact, thus, the suffixes in the last $25\%$ positions of $\Psi$ --- those corresponding to deactivation time encodings --- points to the suffixes corresponding to the source vertex encoding of the same contact.
Hence, an algorithm to answer $neighbors(\mathcal{G}, u, t_{begin}, t_{end})$, first encodes $v$ and, then, it iterates the suffixes that starts with the $v$ encoding to construct the candidate contacts $c = \{ u', v', t_{begin}', t_{end}' \}$, where, in this case, $u' = M(rank_1(D, \Psi[\Psi[\Psi[i]]]) - 1)$, $v' = v$, $t_{begin}' = M(rank_1(D, \Psi[i]) - 1)$ and $t_{end}' = M(rank_1(D, \Psi[\Psi[i]]) - 1)$.
The next steps are the same as in $neighbors(\mathcal{G}, u, t_{begin}, t_{end})$.
Therefore, $neighbors(\mathcal{G}, u, t_{begin}, t_{end})$ and $neighbors(\mathcal{G}, u, t_{begin}, t_{end})$ have the same time complexity.

\subsection{Compressed $\vect{k^d}$ Tree}

\begin{figure}

\begin{adjustwidth}{-2in}{-2in}
    \centering
    \subfloat[]{
\begin{tikzpicture}[font=\ttfamily, square/.style={draw=black!40,outer sep=0pt,minimum width=2em,minimum height=2em}, scale=0.80, every node/.style={transform shape}]
\coordinate (C0@);
\node[square,anchor=west,draw, fill=white] (C00) at (C0@.east) {$0$}
 node[anchor=south, font=\scriptsize\ttfamily] at (C00.north) {$0$};
\node[anchor=east, font=\scriptsize\ttfamily] at (C00.west) {$0$};
\node[square,anchor=west,draw, fill=black!30] (C01) at (C00.east) {$1$}
 node[anchor=south, font=\scriptsize\ttfamily] at (C01.north) {$1$};
\node[square,anchor=west,draw, fill=white] (C02) at (C01.east) {$0$}
 node[anchor=south, font=\scriptsize\ttfamily] at (C02.north) {$2$};
\node[square,anchor=west,draw, fill=black!30] (C03) at (C02.east) {$1$}
 node[anchor=south, font=\scriptsize\ttfamily] at (C03.north) {$3$};
\node[square,anchor=west,draw, fill=white] (C04) at (C03.east) {$0$}
 node[anchor=south, font=\scriptsize\ttfamily] at (C04.north) {$4$};
\node[square,anchor=west,draw, fill=white] (C05) at (C04.east) {$0$}
 node[anchor=south, font=\scriptsize\ttfamily] at (C05.north) {$5$};
\node[square,anchor=west,draw, fill=white] (C06) at (C05.east) {$0$}
 node[anchor=south, font=\scriptsize\ttfamily] at (C06.north) {$6$};
\node[square,anchor=west,draw, fill=white] (C07) at (C06.east) {$0$}
 node[anchor=south, font=\scriptsize\ttfamily] at (C07.north) {$7$};
\node[square,anchor=north,draw, fill=white] (C10) at (C00.south) {$0$};
\node[anchor=east, font=\scriptsize\ttfamily] at (C10.west) {$1$};
\node[square,anchor=north,draw, fill=white] (C11) at (C01.south) {$0$};
\node[square,anchor=north,draw, fill=black!30] (C12) at (C02.south) {$1$};
\node[square,anchor=north,draw, fill=white] (C13) at (C03.south) {$0$};
\node[square,anchor=north,draw, fill=black!30] (C14) at (C04.south) {$1$};
\node[square,anchor=north,draw, fill=white] (C15) at (C05.south) {$0$};
\node[square,anchor=north,draw, fill=white] (C16) at (C06.south) {$0$};
\node[square,anchor=north,draw, fill=white] (C17) at (C07.south) {$0$};
\node[square,anchor=north,draw, fill=white] (C20) at (C10.south) {$0$};
\node[anchor=east, font=\scriptsize\ttfamily] at (C20.west) {$2$};
\node[square,anchor=north,draw, fill=white] (C21) at (C11.south) {$0$};
\node[square,anchor=north,draw, fill=white] (C22) at (C12.south) {$0$};
\node[square,anchor=north,draw, fill=white] (C23) at (C13.south) {$0$};
\node[square,anchor=north,draw, fill=white] (C24) at (C14.south) {$0$};
\node[square,anchor=north,draw, fill=white] (C25) at (C15.south) {$0$};
\node[square,anchor=north,draw, fill=white] (C26) at (C16.south) {$0$};
\node[square,anchor=north,draw, fill=white] (C27) at (C17.south) {$0$};
\node[square,anchor=north,draw, fill=white] (C30) at (C20.south) {$0$};
\node[anchor=east, font=\scriptsize\ttfamily] at (C30.west) {$3$};
\node[square,anchor=north,draw, fill=black!30] (C31) at (C21.south) {$1$};
\node[square,anchor=north,draw, fill=white] (C32) at (C22.south) {$0$};
\node[square,anchor=north,draw, fill=white] (C33) at (C23.south) {$0$};
\node[square,anchor=north,draw, fill=white] (C34) at (C24.south) {$0$};
\node[square,anchor=north,draw, fill=white] (C35) at (C25.south) {$0$};
\node[square,anchor=north,draw, fill=white] (C36) at (C26.south) {$0$};
\node[square,anchor=north,draw, fill=white] (C37) at (C27.south) {$0$};
\node[square,anchor=north,draw, fill=white] (C40) at (C30.south) {$0$};
\node[anchor=east, font=\scriptsize\ttfamily] at (C40.west) {$4$};
\node[square,anchor=north,draw, fill=white] (C41) at (C31.south) {$0$};
\node[square,anchor=north,draw, fill=white] (C42) at (C32.south) {$0$};
\node[square,anchor=north,draw, fill=black!30] (C43) at (C33.south) {$1$};
\node[square,anchor=north,draw, fill=white] (C44) at (C34.south) {$0$};
\node[square,anchor=north,draw, fill=white] (C45) at (C35.south) {$0$};
\node[square,anchor=north,draw, fill=white] (C46) at (C36.south) {$0$};
\node[square,anchor=north,draw, fill=white] (C47) at (C37.south) {$0$};
\node[square,anchor=north,draw, fill=white] (C50) at (C40.south) {$0$};
\node[anchor=east, font=\scriptsize\ttfamily] at (C50.west) {$5$};
\node[square,anchor=north,draw, fill=white] (C51) at (C41.south) {$0$};
\node[square,anchor=north,draw, fill=white] (C52) at (C42.south) {$0$};
\node[square,anchor=north,draw, fill=white] (C53) at (C43.south) {$0$};
\node[square,anchor=north,draw, fill=white] (C54) at (C44.south) {$0$};
\node[square,anchor=north,draw, fill=white] (C55) at (C45.south) {$0$};
\node[square,anchor=north,draw, fill=white] (C56) at (C46.south) {$0$};
\node[square,anchor=north,draw, fill=white] (C57) at (C47.south) {$0$};
\node[square,anchor=north,draw, fill=white] (C60) at (C50.south) {$0$};
\node[anchor=east, font=\scriptsize\ttfamily] at (C60.west) {$6$};
\node[square,anchor=north,draw, fill=white] (C61) at (C51.south) {$0$};
\node[square,anchor=north,draw, fill=white] (C62) at (C52.south) {$0$};
\node[square,anchor=north,draw, fill=white] (C63) at (C53.south) {$0$};
\node[square,anchor=north,draw, fill=white] (C64) at (C54.south) {$0$};
\node[square,anchor=north,draw, fill=white] (C65) at (C55.south) {$0$};
\node[square,anchor=north,draw, fill=white] (C66) at (C56.south) {$0$};
\node[square,anchor=north,draw, fill=white] (C67) at (C57.south) {$0$};
\node[square,anchor=north,draw, fill=white] (C70) at (C60.south) {$0$};
\node[anchor=east, font=\scriptsize\ttfamily] at (C70.west) {$7$};
\node[square,anchor=north,draw, fill=white] (C71) at (C61.south) {$0$};
\node[square,anchor=north,draw, fill=white] (C72) at (C62.south) {$0$};
\node[square,anchor=north,draw, fill=white] (C73) at (C63.south) {$0$};
\node[square,anchor=north,draw, fill=white] (C74) at (C64.south) {$0$};
\node[square,anchor=north,draw, fill=white] (C75) at (C65.south) {$0$};
\node[square,anchor=north,draw, fill=white] (C76) at (C66.south) {$0$};
\node[square,anchor=north,draw, fill=white] (C77) at (C67.south) {$0$};
\draw[line width=1.6, black] ($(C00.north west) + (0, -0)$) rectangle ($(C77.south east) - (0, -0)$);
\draw[line width=0.8, black!80] ($(C00.north west) + (0.008, -0.008)$) rectangle ($(C33.south east) - (0.008, -0.008)$);
\draw[line width=0.4, black!60] ($(C00.north west) + (0.016, -0.016)$) rectangle ($(C11.south east) - (0.016, -0.016)$);
\draw[line width=0.4, black!60] ($(C02.north west) + (0.016, -0.016)$) rectangle ($(C13.south east) - (0.016, -0.016)$);
\draw[line width=0.4, black!60] ($(C20.north west) + (0.016, -0.016)$) rectangle ($(C31.south east) - (0.016, -0.016)$);
\draw[line width=0.4, black!60] ($(C22.north west) + (0.016, -0.016)$) rectangle ($(C33.south east) - (0.016, -0.016)$);
\draw[line width=0.8, black!80] ($(C04.north west) + (0.008, -0.008)$) rectangle ($(C37.south east) - (0.008, -0.008)$);
\draw[line width=0.4, black!60] ($(C04.north west) + (0.016, -0.016)$) rectangle ($(C15.south east) - (0.016, -0.016)$);
\draw[line width=0.4, black!60] ($(C06.north west) + (0.016, -0.016)$) rectangle ($(C17.south east) - (0.016, -0.016)$);
\draw[line width=0.4, black!60] ($(C24.north west) + (0.016, -0.016)$) rectangle ($(C35.south east) - (0.016, -0.016)$);
\draw[line width=0.4, black!60] ($(C26.north west) + (0.016, -0.016)$) rectangle ($(C37.south east) - (0.016, -0.016)$);
\draw[line width=0.8, black!80] ($(C40.north west) + (0.008, -0.008)$) rectangle ($(C73.south east) - (0.008, -0.008)$);
\draw[line width=0.4, black!60] ($(C40.north west) + (0.016, -0.016)$) rectangle ($(C51.south east) - (0.016, -0.016)$);
\draw[line width=0.4, black!60] ($(C42.north west) + (0.016, -0.016)$) rectangle ($(C53.south east) - (0.016, -0.016)$);
\draw[line width=0.4, black!60] ($(C60.north west) + (0.016, -0.016)$) rectangle ($(C71.south east) - (0.016, -0.016)$);
\draw[line width=0.4, black!60] ($(C62.north west) + (0.016, -0.016)$) rectangle ($(C73.south east) - (0.016, -0.016)$);
\draw[line width=0.8, black!80] ($(C44.north west) + (0.008, -0.008)$) rectangle ($(C77.south east) - (0.008, -0.008)$);
\draw[line width=0.4, black!60] ($(C44.north west) + (0.016, -0.016)$) rectangle ($(C55.south east) - (0.016, -0.016)$);
\draw[line width=0.4, black!60] ($(C46.north west) + (0.016, -0.016)$) rectangle ($(C57.south east) - (0.016, -0.016)$);
\draw[line width=0.4, black!60] ($(C64.north west) + (0.016, -0.016)$) rectangle ($(C75.south east) - (0.016, -0.016)$);
\draw[line width=0.4, black!60] ($(C66.north west) + (0.016, -0.016)$) rectangle ($(C77.south east) - (0.016, -0.016)$);
\end{tikzpicture}
} \qquad
    \subfloat[]{
\begin{tikzpicture}[every tree node/.style={font=\ttfamily}, sibling distance=0pt, level distance=40pt, scale=0.70, every node/.style={transform shape}]
\Tree[.~
    [.1
      [.1 0 1 0 0 ]
      [.1 0 1 1 0 ]
      [.1 0 0 0 1 ] 0 ]
    [.1
      [.1 0 0 1 0 ] 0 0 0 ]
    [.1 0
      [.1 0 1 0 0 ] 0 0 ] 0 ];
;\end{tikzpicture}
} \\
    \subfloat[]{
\begin{tikzpicture}[every tree node/.style={font=\ttfamily}, every leaf node/.style={yshift=28pt}, sibling distance=0pt, level distance=40pt, scale=0.90, every node/.style={transform shape}]
\Tree[.~
    [.\node[fill=black!35]{1}; [.\node[fill=black!70]{1}; \edge[draw=none]; \node {\tiny $p_{2}:(1,0)$}; ]
      [.\node[fill=black!35]{1}; [.0 \edge[draw=none]; \node{}; ] [.\node[fill=black!70]{1}; \edge[draw=none]; \node {\tiny $p_{4}:(3,0)$}; ] [.\node[fill=black!70]{1}; \edge[draw=none]; \node {\tiny $p_{5}:(2,1)$}; ] [.0 \edge[draw=none]; \node{}; ] ] [.\node[fill=black!70]{1}; \edge[draw=none]; \node {\tiny $p_{3}:(1,3)$}; ] [.0 \edge[draw=none]; \node{}; ] ] [.\node[fill=black!70]{1}; \edge[draw=none]; \node {\tiny $p_{0}:(4,1)$}; ] [.\node[fill=black!70]{1}; \edge[draw=none]; \node {\tiny $p_{1}:(3,4)$}; ] [.0 \edge[draw=none]; \node{}; ] ];
\end{tikzpicture}
} \qquad
    \subfloat[]{
\begin{tikzpicture}[square/.style={draw,outer sep=0pt,inner sep=1.85pt,text depth=.25ex,text height=1.75ex,font=\scriptsize\ttfamily,minimum width=4.1mm}, scale=0.90, every node/.style={transform shape}]
\node[square] (N0) {1}
 node[square, right] (N1) at (N0.east) {1}
 node[square, right] (N2) at (N1.east) {1}
 node[square, right] (N3) at (N2.east) {0}
 node[square, right] (N4) at (N3.east) {1}
 node[square, right] (N5) at (N4.east) {1}
 node[square, right] (N6) at (N5.east) {1}
 node[square, right] (N7) at (N6.east) {0}
 node[square, right] (N8) at (N7.east) {0}
 node[square, right] (N9) at (N8.east) {1}
 node[square, right] (N10) at (N9.east) {1}
 node[square, right] (N11) at (N10.east) {0};
\node[left= 0.3em of N0] {\normalsize $N$};
\node[square, anchor=north, yshift=-3mm] (B0) at (N0.south) {0}
 node[square, right] (B1) at (B0.east) {1}
 node[square, right] (B2) at (B1.east) {1}
 node[square, right] (B3) at (B2.east) {1}
 node[square, right] (B4) at (B3.east) {0}
 node[square, right] (B5) at (B4.east) {1}
 node[square, right] (B6) at (B5.east) {1}
 node[square, right] (B7) at (B6.east) {1};
\node[left= 0.3em of B0] (B@) {\normalsize $B$};
\node[anchor=north, yshift=-3mm] (C@) at (B@.south) {\normalsize $C$};
\node[square, right=0.3em] (C0) at (C@.east) {$(0, 1)$}
 node[square, right] (C1) at (C0.east) {$(3, 0)$}
 node[square, right] (C2) at (C1.east) {$(1, 0)$}
 node[square, right] (C3) at (C2.east) {$(1, 1)$};
\node[black!90] at ($(C0) - (0, 0.4)$) {\tiny $p_{0}$};
\node[black!90] at ($(C1) - (0, 0.4)$) {\tiny $p_{1}$};
\node[black!90] at ($(C2) - (0, 0.4)$) {\tiny $p_{2}$};
\node[black!90] at ($(C3) - (0, 0.4)$) {\tiny $p_{3}$};
\end{tikzpicture}
}
\end{adjustwidth}
    \caption{$k^d$ tree representation, with $k = 2$ and $d = 2$, of the underlying non-temporal graph shown in Figure~\ref{fig:temporal-graphs} (only edges without timestamps). In~(a), we show the 2-dimensional matrix that stores the binary relations between source and target vertices, $u$ and $v$, respectively.
In~(b), we show the corresponding $k^d$ tree without path compression.
Each node has $k^d$ children representing $k^d$ different partitions of size $m / 2^{l - 1}$ inside the original matrix, where $l$ is the level of a node in the tree.
In~(c), we show the corresponding c$k^d$ tree, which compresses whole paths into black nodes.
Instead of storing whole paths that results in single bits set to $1$, the c$k^d$ tree stores only the global coordinates regarding these single bits in a different type of node.
Finally, in~(d), we show the memory layout of the c$k^d$ tree that contains a bitarray $N$, with information about bits set, a bitarray $B$, with information about colors for bits set to $1$ and an array $C$ of relative coordinates from the current partitions.
Note that $C$ does not store global coordinates to save space.
Also, it does not store positions of black nodes in the last level since these positions can be obtained during searches.}\label{fig:k2tree}
\end{figure}
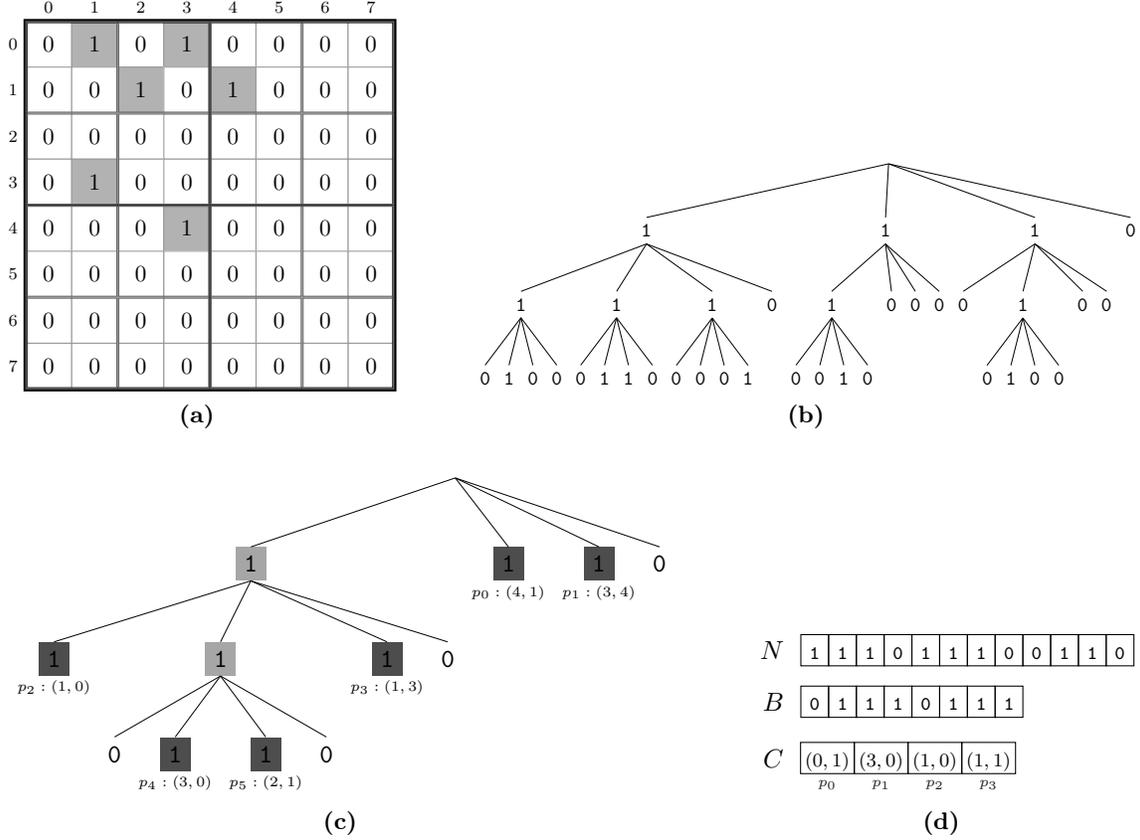

Compressed $k^d$ tree (c$k^d$-tree)~\citep{caro2016compressed} is a compressed version of the $k^d$-tree that stores $d$-dimensional bitmaps efficiently~\citep{de2014new}\footnote{Source code is available at \url{https://github.com/diegocaro/tgdim/}.}.
As shown in Figure~\ref{fig:k2tree}, the c$k^d$-tree represents recursively the decomposition of a $d$-dimensional bitmap into equal sized partitions.
At each level, it splits the current bitmap partitions of size $(s_1, s_2, \ldots, s_d)$ into $k^d$ smaller partitions of size $(\frac{s_1}{k}, \frac{s_2}{k}, \ldots, \frac{s_d}{k})$ and redirects them to the lower level nodes.
Each node stores a $1$-dimensional bitmap $B$ of size $k^d$ to describe which partition can be further split.
If the current partition only contains bits $0$ then there is no need to split it further and, therefore, the corresponding position at bitmap $B$ is set to $0$.
Otherwise, if it contains some bit $1$ then the corresponding position at bitmap $B$ is set to $1$ and, additionally, the node holds a pointer to the next child that will split it further.
To check the state of some bit $b$ in a $d$-dimensional bitmap at position $p = (p_1, p_2, \ldots, p_d)$, an algorithm traverses the tree following a top-down approach and, at each level, it search for the $i$-th partition that contains $p$ and descend to the corresponding child whether $B_i = 1$.
If at some point $B_i = 0$ then $b = 0$, otherwise, if the algorithm reaches an external node with $B_i = 1$ then $b = 1$.

\cite{caro2016compressed} stores and query temporal graphs by using c$k^4$-trees with each dimension representing, respectively, source vertices, target vertices, activation times and deactivation times of contacts.
As temporal graphs are usually sparse, an naive implementation of the c$k^4$-tree structure would traverse many nodes with only one child until it reaches an external node.
In order to decrease the number of nodes with only one child and, consequently, improve space and query efficiency, the c$k^4$-tree uses a second type of external node to store only the relative coordinates of the cell that has the single value $1$ inside the current partition.
Therefore, internal nodes also must store an additional $1$-dimensional bitmap to differentiate the type of external nodes, with values being $0$ if the corresponding children represent partitions with more than one bit $1$ or $1$ if children represent partitions with a single bit set to $1$.

\subsubsection{Operation $\vect{has\mathunderscore{}edge}$ in c$\vect{k^d}$-tree}

An algorithm to answer \changed{\hasedgei} uses the $point(\mathcal{T}, P)$ query internally, passing the c$k^4$-tree $\mathcal{T}$ constructed from $\mathcal{G}$ and the $d$-dimensional point of interest $P = \langle u, v, t_{begin}, t_{end} \rangle$.
Starting at the root node, the $point(\mathcal{T}, P)$ query algorithm descends the c$k^d$-tree recursively checking at every level if there is a child partition that may contain $P$.
If the algorithm does not reach an external node then there is no edge $(u, v)$ active during the interval $[t_{begin}, t_{end}]$.
Otherwise, depending on the type of the external node, it verifies whether the cell that contains $P$ in the current partition is set to $1$ to check if an edge $(u, v)$ is active during the interval $[t_{begin}, t_{end}]$ or not.
If the external node has an $1$-dimensional bitmap $B$ (type 1), the algorithm calculates the relative position $P_r$ in the current partition using the path it traversed to reach this node and, then, it checks whether $B_i = 1$, where $i$ is the position associated with the cell $P_r$ in the current partition.
Otherwise, if the node stores only the relative position $C_r$ of the single bit $1$ inside the corresponding partition (type 2), the algorithm calculates $P_r$ and, then, it checks whether $P_r = P_r$.

\subsubsection{Operation $\vect{neighbors}$ in c$\vect{k^d}$-tree}

An algorithm to answer \changed{\neighsi} uses the $range(\mathcal{T}, R)$ query internally by passing the c$k^4$-tree $\mathcal{T}$ constructed from $\mathcal{G}$ and the $d$-dimensional region of interest $R$ formed by the lower boundary $L = \langle u, \min{(V)}, \min{(T)}, t_{begin} + 1 \rangle$ and the upper boundary $U = \langle u, \max{(V)}, t_{begin}, \max{(T)} \rangle$.
Starting at the root node, the $range(\mathcal{T}, U)$ query algorithm descends recursively all children which partitions overlap $R$ until it reaches all possible external nodes.
Then, for each external node, the algorithm computes the bit $1$ coordinates that overlap $R$ depending on its type and adds them to the result set.
If the external node has a $1$-dimensional bitmap $B$ (type 1), the algorithm searches for the positions $i$ where $B_i = 1$, compute the relative positions associated with positions $i$ and calculate the global coordinates based on the path it traversed to reach the node.
Otherwise, if the node stores only the relative position of the single bit $1$ inside the corresponding partition (type 2), the algorithm simply calculates the global coordinate based on the path it traversed.

\subsubsection{Operation $\vect{neighbors^r}$ in c$\vect{k^d}$-tree}

An algorithm to process \changed{\rneighsi} is similar to the direct query.
However, it fixes the second dimension --- the dimension associated with target vertices --- instead of the first when calling $range(\mathcal{T}, R)$, thus, the region $R$ is formed by $L = (\min{(V)}, v, \min{(T)}, t_{begin} + 1)$ and $U = (\max{(V)}, v, t_{begin}, \max{(T)})$.
As CET, finding direct and inverse neighbors using TGCSA have the same time complexity.

\section{Discussion}\label{sec:discussion}

In this section we compare the data structures we review in Section~\ref{sec:primary-memory} based on their worst-case space cost and time cost for answering some queries we described in Section~\ref{sec:background}.
In order to get the information of costs, we simplified the expanded formula, when available, in the original using the big-$O$ notation.
The considered variables are: $n$, for the number of vertices; $m$, for the number of edges in the underlying static graph; $c$, for the number of contacts; and $t$, the for lifetime of the temporal graph.
For a more detailed description see the work by~\cite{caro2015data}.

\begin{table}[]
    \centering
    \caption{Worst-case space cost of the temporal graph structures using the number of vertices $n$, number of edges $m$, number of contacts $c$ and the lifetime $t$ of $\mathcal{G}$.}\label{tab:space-score}
    \begin{tabular}{l|c} \toprule
        \bf Structure & \bf Worst-case Space \\ \midrule
        EveLog & $O(c\log{\frac{n t}{c}} + n\log{(n + c)})$ \\
        EdgeLog & $O(m\log{\frac{n c }{m}} + c\log{\frac{mt}{c}} + n\log{m})$ \\
        CAS & $O(c\log{(n + t)} + n)$  \\
        CET & $O(c\log{m} + t)$  \\
        TGCSA & $O(c \log{(n + t)})$  \\
        c$k^d$ tree ($d=4$) & $O(c \log{\frac{n t}{c}})$  \\ \bottomrule
    \end{tabular}
\end{table}

Table~\ref{tab:space-score} shows the space cost of the data structures we reviewed in Section~\ref{sec:primary-memory}.
In this comparison, we considered the cost of pointers when necessary.
For instance, the structures EveLog and EdgeLog stores pointers to map source vertices and their corresponding temporal adjacency lists or event lists.
The main sources of space consuming for the EveLog structure are the number of contacts and vertices.
This is due to the number of items in the temporal adjacency lists, $c$, and the number of pointers mapping vertices to their corresponding temporal adjacency lists, $n$.

In the case of EdgeLog, there are three major sources of space usage: number of edges, number of contacts and number of vertices.
This is because EdgeLog extends adjacency list for temporal graphs, however, additionally, it stores pointers for time intervals for every edge in the underlying edge, it stores pointers for every vertex to their corresponding temporal adjacency list, and it stores an aditional list of time intervals.

For CAS the major sources of space consumption are the amount of contacts and the amount of vertices.
This is due the size of the sequence to store edge activation and deactivation events which depends $c$ and stores $n + t$ symbols.
Additionally, there is also a bitvector to store positions in which events corresponding to a source vertex begin in the sequence based on the amount of vertices $n$.

CET increases space consumption according to the amount of contacts and the lifetime of the temporal graph.
Similarly to CAS, CET stores a sequence with size based on the amount of contacts, however, the number of symbols is based on the number of the underlying edges and the size of the bitvector is based on the variable $t$ since it marks the start of each timestamp in the sequence.

The major source of space consumption of TGCSA is linked to the amount of contacts since it compresses the sequence containing the concatenation of all contacts and the suffix array of the same size containing the surroundings of each symbol in the original sequence.
This is due that TGCSA store a constant amount of data structures with size depending on the amount of contacts.
For example, it needs to store a dictionary that maps every symbol in contacts to their corresponding code, a bitvector that stores information about symbols in the ordered suffix array and a sequence that stores information about the surroundings of symbols in the original sequence.

In the case of c$k^d$ tree, being $d = 4$ due to the dimension of contacts, the major source of space consumption is the number of contacts.
This is because the c$k^4$ tree compresses a tensor of degree $4$ containing $c$ bits set with dimensions size based on the number of vertices $n$ and the lifetime of the temporal graph $t$.

\begin{table}[]
    \caption{Time cost of queries with a timestamp parameter using the number of vertices $n$, number of edges $m$, number of contacts $c$ and the lifetime $t$ of $\mathcal{G}$.}\label{tab:queries-score}
\begin{adjustwidth}{-2in}{-2in}
    \centering
    \begin{tabular}{l|c|c|c|c} \toprule
      \bf Structure & \bf \changed{\hasedgename} & \bf \changed{\neighsname} & \bf \changed{\rneighsname} & \bf \changed{\aggregname} \\ \midrule
        EveLog & $O(\frac{c}{n})$ & $O(\frac{c + m}{n})$ & infeasible & $O(c + m)$ \\
        EdgeLog & $O(\frac{c}{m} + \frac{m}{n} + \log{(\frac{c}{n})})$ & $O(\frac{c}{n} + \frac{m}{n}\log{(\frac{c}{m})})$ & infeasible &  $O(c + m\log{(\frac{c}{m})})$ \\
        CAS & $O(\log{(n + t)})$ & $O(\frac{m}{n} \log{(n + t)})$ & $O(\frac{c + m}{n}\log{(n + t)})$ & $O(m \log{(n + t)})$  \\
        CET & $O(\log{n})$  & $O(\frac{m}{n} \log{n})$ & $O(\frac{m}{n} \log{n})$ & $O(m \log{n})$  \\
        TGCSA & $O(\frac{c}{m} \log{(c)})$ & $O(\frac{c}{n}\log{(c)})$ & $O(\frac{c}{n}\log{(c)})$ & $O(c\log{c})$  \\
        c$k^d$ tree ($d = 4$) & $O(c^{\frac{1}{2}})$ & $O(c^{\frac{3}{4}})$ & $O(c^{\frac{3}{4}})$ & $O(c)$ \\ \bottomrule
        \multicolumn{5}{c}{- Uniform degree distribution in the aggregate graph was considered when necessary} \\
    \end{tabular}
\end{adjustwidth}
\end{table}

Table~\ref{tab:queries-score} shows the time cost of the data structures to answer some queries we described in Section~\ref{sec:background}.
In this table, we compare the following queries: \changed{\hasedgename}, \changed{\neighsname}, \changed{\rneighsname} and \changed{\aggregname}.
We note that there are only queries based on a single timestamp and for the costs, as the original authors, we considered a graph generated using uniform degree distribution.
As we can see, for EveLog, the cost of the query \changed{\hasedgename} depends on the average number of events of activation or deactivation in the temporal adjacency list associated with vertex $u$ to find the events with vertex $v$ until time $t$.
Similarly, the cost of \changed{\neighsname} also depends on the average number of events of the source vertex $u$ but it needs also to consider the average number of edges in the underlying graph in that $u$ participates.
The \changed{\rneighsname} is infeasible using only the temporal adjacency lists for the outgoing contacts because it would traverse the entire structure.
As suggested in Section~\ref{sec:primary-memory}, one can store another structure that considers the incoming contacts of a vertex and get time costs similar with the \changed{\neighsname} query, however, it would double the space needed.
Finally, for \changed{\aggregname}, EveLog needs to computes one \changed{\neighsname} for all $u \in V$ to construct the snapshot $\mathcal{G}_t$.

For EdgeLog, the query \changed{\hasedgename} needs to decompress the temporal adjacency list associated with $u$ with average size $\frac{m}{n}$ and the list of time intervals associated with edge $(u, v)$ with size $\frac{c}{m}$.
Also, it needs to run binary searches to check if this edge is active at timestamp $t$ using a binary search.
For the query \changed{\neighsname}, EdgeLog need to decompress the temporal adjacency list associated with vertex $u$ and all lists of time intervals of edges $(u, v)$.
Then it needs to binary search the lists containing intervals to check if edge $(u, v)$ is active at time $t$.
For the same reasoning of EveLog, the query \changed{\rneighsname} is infeasible for EdgeLog and it also can spend about the double the space to have similar costs of \changed{\neighsname}.
Finally, \changed{\aggregname} uses one \changed{\neighsname} for every $u \in V$ and a binary search is performed for every edge $(u, v)$ in the corresponding list of time intervals. During this process, all the temporal graph ends up being decompressed.

Differently, the other data structures do not need a decompressing step.
For CAS, the query \changed{\hasedgename} perform operations in the underlying wavelet matrix that stores a sequence representing temporal adjacency lists in for form of events of activation and deactivation.
As this sequence has $n + t$ symbols, a query to count the amount of $(v, t)$ occurrences in the block associated with $v$ is $O(\log({n + t}))$.
For query \changed{\neighsname}, this same reasoning is made for every $(u, v)$.
In this case, the average value per vertex is $\frac{m}{n}$ and, therefore, the total cost is $O(\frac{m}{m}\log{(n + t)})$.
The \changed{\rneighsname} query becomes feasible with the CAS structure because it does not need to decompress the whole structure as the other two structures.
However, it still needs to call one wavelet tree operation in every block associated with some other vertex $u \in V$ to count the number of events $(u, v)$ in time $t$.
Finally, for \changed{\aggregname}, CAS calls the query \changed{\neighsname} for every $u \in V$ and, thus, it needs to execute one wavelet matrix operation for each edge $(u, v)$ for $u,v \in V \times V$.

For CET, the query \changed{\hasedgename} also uses the operation to count occurrences on the underlying wavelet matrix representing the sequence of events of activation or deactivation.
However, different of CAS, it stores on the sequence pair of vertices or edges $(u, v)$ and uses the additional bitvector to split the sequence in blocks with the same timestamp.
Therefore, in the sequence there are only vertex symbols and, therefore, the cost to count the number of occurrences of $(u, v)$ in the block associated with $t$ is $O(\log{n})$.
For the query \changed{\neighsname}, CET calls $\frac{m}{n}$ times, the average number of edges for vertex, the wavelet tree operation, thus the total cost is  $O(\frac{m}{n}\log{n})$.
The algorithm for \changed{\rneighsname} is similar to \changed{\neighsname}, it only needs to change the symbols $(u', v)$ being counted and, thus, CET is the first compact data structure to present the same costs for the both operations.
Finally, for \changed{\aggregname}, similar to the other structures so far, it needs to call \changed{\neighsname} for all $u \in V$ and, therefore, it counts occurrences for all $m$ edges $(u, v)$ with $v \in V$ also.

For TGCSA, the query \changed{\hasedgename} spents $O(\log{c})$ cost to convert symbols from the data structures to symbols in the original sequence of concatenated contacts.
Therefore, after each call to the array $\Psi$, the TGCSA needs to decode the symbol using the dictionary to reason about the resulting symbol.
As the substring $uv$ can discard mostly candidates in the string matching algorithm, there are the average $O(\frac{c}{m})$ timestamps to be checked for edge $u, v$.
For \changed{\neighsname}, the algorithm cannot filter contact candidates in the string matching process as for \changed{\hasedgename} using symbol $v$, therefore it calls $O(\frac{c}{n})$ times the access operation in the array $\Psi$ and, thus, the same amount of operations for decoding the resulting symbol using the dictionary.
Similarly to CET, the TGCSA structure also can answer the query \changed{\rneighsname} with the same cost of the $neighbors(\mathcal{G}, u, t)$ query.
The reason is that the subsequence matching can be circular using the array $\Psi$ and, thus, the process can start at symbol filtering candidates by the coded symbol for symbol $t$ and then use $u$ to continue the process.
Finally, for \changed{\aggregname} TGCSA also needs to call $n$ times the query \changed{\neighsname} for all $u \in V$ and, therefore, all contact symbols must be accessed and decoded.

For the c$k^d$ tree, all the graph queries are translated to a range query in the $4$-dimensional tensor using a $4$-dimensional rectangle $R = (L, U)$ consisting of the lower and upper bound points $L$ and $U$, respectively.
For the query \changed{\hasedgename}, as there are $2$ dimensions known, the algorithm constructs an rectangle $R = (L, U)$, where $L = \langle u, v, 0, t \rangle$ and $U = \langle u, v, t, \max(T) \rangle$.
Note that the algorithm needs to search in the region where $t_{begin} \leq t$ and $t_{end} \geq t$ to find contacts with intervals that contains $t$.
Therefore, at every level of the tree, the search algorithms can descend to $k^2$ children since half associated with the two first dimensions do not pass the test.
Also, the height of the three is $h = \log_{k^4}{c} = \frac{1}{4}\log_k{c}$ and, thus, the algorithm visits at most ${(k^2)}^h = O(c^\frac{2}{4})$ nodes.
For the query \changed{\neighsname}, there are only one known dimension and the algorithm constructs the rectangle consisting of $L = \langle u, 0, 0, t \rangle$ and $U = \langle u, \max(V), t, \max(T) \rangle$.
Therefore, the algorithm descends at most ${(k^3)}^h = O(c^\frac{3}{4})$ nodes since just $k^1$ nodes are not visited at every level.
For the query \changed{\rneighsname}, the c$k^d$ tree has the same cost as there is only one fixed dimension and the constructed rectangle consists of  $L = \langle 0, v, 0, t \rangle$ and $U = \langle \max(V), v, t, \max(T) \rangle$.
Finally, for the query \changed{\aggregname} there is no fixed dimension.
Therefore, it constructs a rectangle consisting of $L = \langle 0, 0, 0, t \rangle$ and $U = \langle \max(V), \max(V), t, \max(T) \rangle$ and descends all $k^4$ children every level and, thus, the cost is ${(k^4)}^h = O(c)$.

\section{Conclusion}\label{sec:conclusion}

This paper reviewed studies about data structures for storing and querying temporal graphs in primary memory.
We noted that many authors use data compression techniques to reduce the amount of space per contact needed to store temporal graphs.
There are also succinct data structures that allows a variety of useful queries with time complexity of queries similar to non-compressed data structures.
Therefore, algorithms could efficiently process large amount of data in primary memory.

For future works, we suggest the development of techniques that circumvents the weaknesses of the discussed data structures.
For example, EdgeLog can perform direct neighbor queries efficiently, however efficient reverse neighbor queries need duplication of storage.
CET and c$k^2$ tree have the same complexity for direct and reverse neighbor queries, however, accessing a contact has logarithmic complexity.
TGCSA can also perform direct and reverse neighbor queries, however, it has a filtering step for each dimension, even for more basic queries such as checking whether a contact is active at some timestamp.

\section*{Acknowledgements}

This study was financed in part by Funda\c{c}\~{a}o de Amparo \`{a} Pesquisa do Estado de Minas Gerais (FAPEMIG) and the Coordena\c{c}\~{a}o de Aperfei\c{c}oamento de Pessoal de N\'{i}vel Superior - Brasil (CAPES) - Finance Code 001* - under the ``CAPES PrInt program'' awarded to the Computer Science Post-graduate Program of the Federal University of Uberl\^{a}ndia.

\printbibliography{}

\end{document}